\documentclass[lettersize,journal]{IEEEtran}
\usepackage{amsmath,amsfonts}
\usepackage{array}
\usepackage{textcomp}
\usepackage{stfloats}
\usepackage{url}
\usepackage{verbatim}
\usepackage{cite}
\usepackage{enumerate}
\usepackage{graphicx}
\usepackage{amssymb}
\usepackage{amsfonts}
\usepackage{mathtools}
\usepackage{xspace}
\usepackage{subcaption}
\usepackage[lined,linesnumbered,ruled,commentsnumbered]{algorithm2e} 
\usepackage{colonequals}
\usepackage{fixltx2e}    
\usepackage{amsthm}
\usepackage{hyperref}
\usepackage{caption}
\usepackage{float}
\usepackage{relsize}
\usepackage{ftnxtra}
\usepackage{fnpos} 

\usepackage{tikz}
\usepackage{pgfplots}
\pgfplotsset{scaled y ticks=false}
\usetikzlibrary{arrows,shapes,backgrounds,plotmarks,positioning}
\pgfplotsset{compat=newest}                         
\pgfplotsset{plot coordinates/math parser=false}
\newlength\figureheight
\newlength\figurewidth

\newtheorem{theorem}{Theorem}

\newtheorem{definition}{Definition}
\newtheorem{corollary}{Corollary}

\newtheorem{remark}{Remark}

\newcommand{\E}{\mathbb{E}}

\newcommand*\dif{\mathop{}\!\mathrm{d}}

\newcommand{\SA}{\mathcal{S}}

\newcommand{\K}{\mathcal{K}}

\newcommand{\Lap}{\mathcal{L}}
\newcommand{\Gauss}{\mathcal{G}}
\newcommand{\GaussMix}{\mathcal{GM}}

\newcommand{\GaussProb}{\mathfrak{g}}

\newcommand{\Real}{\mathbb{R}}
\newcommand{\RealP}{\mathbb{R}_+}

\newcommand{\Set}[1]{\{#1\}}

\newcommand{\SetPair}{\mathbb{S}}

\begin{document}

\title{Approximation of Pufferfish Privacy for Gaussian Priors}

\author{Ni Ding,~\IEEEmembership{Member,~IEEE}
}

\markboth{Journal of \LaTeX\ Class Files,~Vol.~14, No.~8, August~2021}%
{Shell \MakeLowercase{\textit{et al.}}: A Sample Article Using IEEEtran.cls for IEEE Journals}


\maketitle

\begin{abstract}
This paper studies how to approximate pufferfish privacy when the adversary's prior belief of the published data is Gaussian distributed. Using Monge's optimal transport plan, we show that $(\epsilon, \delta)$-pufferfish privacy is attained if the additive Laplace noise is calibrated to the differences in mean and variance of the Gaussian distributions conditioned on every discriminative secret pair.
A typical application is the private release of the summation (or average) query, for which sufficient conditions are derived for approximating $\epsilon$-statistical indistinguishability in individual's sensitive data.
The result is then extended to arbitrary prior beliefs trained by Gaussian mixture models (GMMs): calibrating Laplace noise to a convex combination of differences in mean and variance between Gaussian components attains $(\epsilon,\delta)$-pufferfish privacy.
\end{abstract}


\section{Introduction}
\IEEEPARstart{W}{hen}
participating in data sharing activities, we want to provide useful information to others but keep secret our personal data, e.g., race, gender, etc.
To do so, some privacy metric is applied to quantify the confidentiality of the sensitive attributes in the released data. A data regulation scheme is devised thereafter to mitigate the privacy leakage.
Differential privacy (DP) \cite{Dwork2006,CalibNoiseDP} is a rigorous definition of data privacy based on a typical inference setting \cite{Dwork2008Survey}.
For an analyst who is able to compute the data statistics,  DP ensures a restricted probabilistic resolution on the secret that is nested in the released data.

Specifically, for a (deterministic) query function $f(\cdot)$ that returns distinct values when it is applied to a database $D$ given two secrets $s_i$ and $s_j$. For example, $s_i$ refers to the event that some user is present in $D$, while $s_j$ denotes the event that this user is absent.
To protect privacy, the query answer $f(D)$ should be randomized (e.g., by injecting noise) to ensure some statistical indistinguishability between $s_i$ and $s_j$. That is, denoting $\Pr(\tilde{f}(D)|s_i)$ and $\Pr(\tilde{f}(D)|s_j)$ the probability of noised $f(D)$ given $s_i$ and $s_j$, respectively, the difference between them should be upper bounded by a nonnegative threshold $\epsilon$.
This is so-called $\epsilon$-indistinguishability, where the \emph{privacy budget} $\epsilon$ denotes the privacy level: a smaller $\epsilon$ indicates higher indistinguishability between $\Pr(\tilde{f}(D)|s_i)$ and  $\Pr(\tilde{f}(D)|s_j) $ and therefore more privacy.\footnote{In DP, it is assumed that the database $D$ given $s_i$ and $s_j$ differs in one entry only.}

While in DP the uncertainty is introduced by the data regulation scheme only, a more realistic scenario is that the original dataset exhibits an inherent uncertainty, too.
For example, a database could be drawn from a probabilistic space, where the chances for getting each realization $D$ conditioned on distinct secrets $s_i$ and $s_j$ are different; or, the query function $f(\cdot)$ could be a randomized response function.
To this end, a more general framework is formulated by \cite{Pufferfish2012KiferConf,Pufferfish2014Kifer} called \emph{pufferfish privacy}.

For the original data $X$ that is statistically correlated with the nesting secret $S$,
the purpose of pufferfish privacy is to have the noised data $\tilde{X}$ probabilistically indistinguishable on $S$. DP can be treated as a special case of pufferfish privacy for deterministic $X$ such that  $X = f(D)$.
This framework also incorporates Bayesian inference setting (as seen in quantitative information flow \cite{QIF2009} and information leakage studies \cite{PvsInfer2012,Lalitha2013TIFS})
via a parameter $\rho$ that denotes the prior belief of the adversary. The prior belief is usually a probability distribution, e.g., the conditional probability of $X$ given secret $S$, before privatization, which could be learned from the previous data release.
In this sense, the posterior probability refers to the statistics of the noised data $\tilde{X}$, where the purpose is to guarantee $\epsilon$-indistinguishability between $\Pr(\tilde{X}|s_i, \rho)$ and $\Pr(\tilde{X}|s_j, \rho)$.

The difficulty is how to calibrate the noise given the intrinsic uncertainty in the original dataset.
It is shown in \cite{PufferfishWasserstein2017Song} that scaling Laplace noise  by the Wasserstein metric of infinite order $W_\infty$ is sufficient to attain $\epsilon$-pufferfish privacy.\footnote{This method reduces to the famous $\ell_1$-sensitivity noise calibration for DP \cite{CalibNoiseDP} if $X$ is deterministic, which also proves that less noise is required for attaining pufferfish privacy than DP~\cite{Pufferfish2014Kifer,PufferfishWasserstein2017Song}. }
\cite{Ding2022Kantorovich} pointed out the infeasibility of this approach due to the non-convexity of  $W_{\infty}$~\cite{Champion2008InfW,DePascale2019InfW} and proposed a realistic $W_1$ noise calibrating method by the corresponding Kantorovich optimal transport plan.
However, the $W_1$ method can only be applied to $X$ taking values in a countable alphabet, as it involves the computation of the second derivative of a joint probability, which is hard to obtain for continuous random variables or probability distributions that do not have a closed-form expression.
In addition, for continuous $X$ and the corresponding probability density function, the maximum pairwise distance over the Kantorovich optimal transport plan could be infinitely large, which would result in an excessive amount of noise and therefore severely deteriorate the data utility.

On the other hand, an adversary would be very likely to adopt machine learning techniques to infer the prior knowledge $\rho$, i.e., train or fit a parameterized probability density function  out of the past observations.
In particular, for $X$ being aggregated statistics, e.g., the counting or summation query, Gaussian prior would be a good choice as it is closest to the true statistics.
This is the reason why the prior knowledge is usually modeled by a normal probability density function such as \cite{Zhang2022AttrPriv}.

\subsection{Our Contributions}
This paper studies how to calibrate Laplace noise for attaining pufferfish privacy when the adversary's prior belief $\rho$ is Gaussian distributed.
The result is further extended to arbitrarily distributed prior belief $\rho$ that is trained by Gaussian mixture model (GMM).
The main results in this paper will be derived under the Monge's optimal transport plan~\cite{Dowson1982Frechet,Givens1984NormClass,Takatsu2011Wnorm}, the $W_2$ solution for Gaussian couplings.

The main contributions of this paper are listed below.

\begin{enumerate}
	\item
	To release data $X$ which is known to be normally distributed given all instances of secret $S$ but differ in mean and variance, we apply Monge's optimal transport plan to show that a $\delta$-approximation of $\epsilon$-pufferfish privacy, i.e., $(\epsilon, \delta)$-pufferfish privacy, can be achieved by adding Laplace noise to $X$. The scale parameter $b$ of Laplace noise should be calibrated to the differences in both mean and variance conditioned on each pair of secrets $s_i$ and $s_j$.
This method is shown to be a generalization of the $\ell_1$-sensitivity noise calibration method for DP \cite{CalibNoiseDP}.

	\item The result above is applied to the problem of privatizing the summation query in a multi-user system containing a finite number of participants. It is assumed that each user obtains an independent random variable.
    To privately release the summation over all users, we derive a sufficient condition for ensuring the statistical indistinguishability about the individual's presence in the system.
	It is proved that $\epsilon$-indistinguishability about each participant's data can be guaranteed.

	\item Assuming the adversary learns the prior knowledge $\rho$ on the arbitrarily distributed $X$ by GMMs, we show that $(\epsilon, \delta)$-pufferfish privacy is guaranteed by calibrating the scale parameter $b$ of Laplace noise to the convex combination of differences in mean and variance of Gaussian components.
We apply this method to the \texttt{adult} and \texttt{Hungarian heart disease} datasets in the UCI machining learning repository \cite{UCI2007} to show how to scale the parameter $b$ to achieves $\delta$-approximation of $\epsilon$-indistinguishability on real-world data.

\end{enumerate}

\subsection{Related Works}

Consider publishing a table having two columns ``\texttt{age}" and ``\texttt{cholesterol level}". Even if column ``\texttt{age}" is excluded, it can still be inferred by an adversary who can exploit the correlation between the two attributes. Here, the DP setting \cite{Dwork2006,CalibNoiseDP} does not fit, as the published data ``\texttt{cholesterol level}" is a random variable, not a deterministic query answer, the statistics of which depend on the hidden secret ``\texttt{age}". Instead, a pufferfish privacy framework is proposed in \cite{Pufferfish2012KiferConf,Pufferfish2014Kifer} to study how to achieve indistinguishability in the presence of intrinsic randomness.
It has been applied to temporally correlated data, e.g., the privacy measure in~\cite{Niu2019_CONF,Ding2022_CONF}, monitoring web browsing behavior~\cite{Liang2020Web} and the trajectory data, e.g.,~\cite{Ou2018_JOURNAL}, etc.
Besides these specific applications, an efficient noise calibrating method remains missing until the proposal of Wasserstein approach in~\cite{PufferfishWasserstein2017Song}.
It is shown that pufferfish privacy can be attained by a Laplace mechanism calibrated by the Wasserstein distance $W_\infty$.
To deal with the difficulty in calculating $W_\infty$, apart from the relaxation by  a R{\'{e}}nyi measure \cite{Pierquin2023renyiPP},  \cite{Ding2022Kantorovich} shows that $W_\infty$ method can be computed by Kantorovich optimal transport plan (the solution to $W_1$ distance), which is easy to obtain for finite and discrete alphabet cases, e.g., the $W_1$ approach for a trajectory clustering task in \cite[Algorithm~2]{Zeng2021}.  This motivates the study of a more realistic case as to how an adversary infers the intrinsic randomness.

Pufferfish privacy is a guarantee of indistinguishability against the adversary's prior knowledge $\rho$~\cite{Pufferfish2012KiferConf,Pufferfish2014Kifer}, the  belief or side information on the statistical dependence of published data on hidden secret, which might be obtained from previous data releases.
For $50$ distinct cholesterol levels, an adversary needs to store $50$ probability values to express the prior belief for only one secret instance.
Instead, fitting a probabilistic model can largely reduce space complexity, e.g., two values, mean and variance, determine a Gaussian probability.
This coincides with the idea of statistical machine learning~\cite{MLbook2015}, but in return causes a problem in the Wasserstein approach: the noise calibrated by the maximum pairwise distance of $W_\infty$ or $W_1$ solution might be too large.\footnote{This is largely due to the fact that the support of a parameterized probability is usually the overall real number set.}
Therefore, the existing study is restricted to special cases of parameterized priors, e.g., Gaussian priors that only differ in mean~\cite{Chen2022AttrPriv,Zhang2022AttrPriv}.
This paper considers pufferfish privacy where the prior $\rho$ is Gaussian distribution with the mean and variance varying with secret instance.
The underlying assumption is that the published data is a continuous random variable, not the bounded discrete ones as in  \cite{Niu2019_CONF,Ding2022_CONF,Liang2020Web,Ou2018_JOURNAL}.

Meanwhile,  an important use case for data privacy is reporting aggregated statistics for a finite number of users/participants, e.g., how to ensure $\epsilon$-DP in counting query so that an adversary cannot easily tell the existence of individual users~\cite{Dwork2006,CalibNoiseDP}.
This also motivates the proposal of pufferfish privacy (see hedging privacy in~\cite[Section~7]{Pufferfish2014Kifer}) where the data obtained by each user is randomized.
This paper also studies how to attain $(\epsilon,\delta)$-pufferfish privacy in this multi-user system.

\subsection{Notation}
The capital and lower case letters denote random variable and its realization, respectively. For example, $x$ denotes an instance/realization of random variable $X$. Notation $P_{X}(x)$ refers to the probability $\Pr(X = x)$.
The calligraphic $\mathcal{X}$ denotes the alphabet of $X$.
We use $X|s$ to denote the random variable $X$ conditioned on the event $S=s$ and $P_{X|S}(\cdot|s)$ to denote the corresponding conditional probability distribution.
Normal probability density distribution with mean $\mu$ and variance $\sigma$ is denoted by $\Gauss(\mu, \sigma^2)$ and Laplace distribution with the scale parameter $b$ is denoted by $\Lap(b)$. We only consider zero mean Laplace distribution in this paper.
We use $\Real$ and $\RealP$ to denote real number set and nonnegative real number set, respectively.

\subsection{Organization}

Section~\ref{sec:Prelim} clarifies the definition of pufferfish privacy and reviews the Monge's optimal transport plan.
Sections~\ref{sec:LapGaussian} and \ref{sec:LapGMM} derive sufficient conditions for attaining $(\epsilon, \delta)$-pufferfish privacy for Gaussian and GMM priors, respectively.
Section~\ref{sec:CountQ} shows how to compute the scaling parameter $b$ for publishing pufferfish private summation query in a multi-user system.
Section~\ref{sec:UCIexp} presents the experimental result on the \texttt{adult} dataset.
In this paper, the main results (Theorems~\ref{theo:Lap} and \ref{theo:LapGaussMix} and Corollaries~\ref{coro:TransPrior} and \ref{coro:TransPriorGMM}) are stated in the main context with corresponding proofs presented in the appendix.

\begin{figure*}[h]
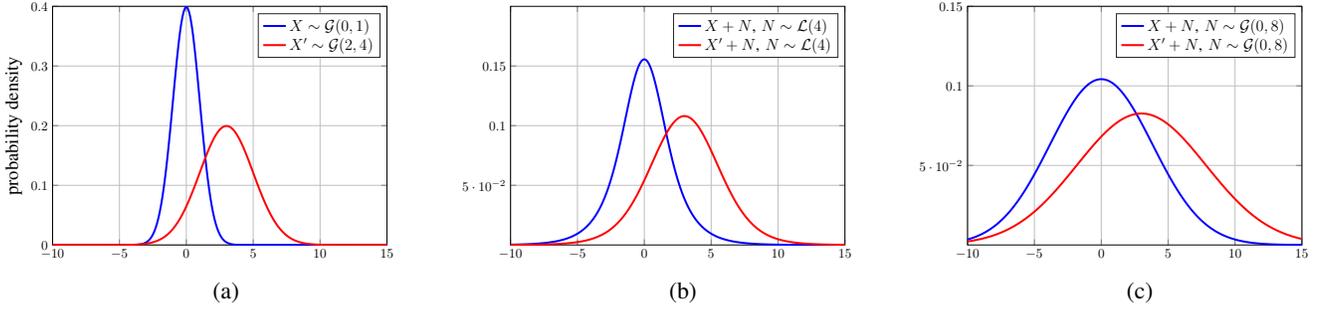

    \begin{subfigure}{0.33\linewidth}
    \scalebox{0.5}{\input{figures/GaussTwo.tex}}
    \caption{}
    \end{subfigure}
    \begin{subfigure}{0.33\linewidth}
    \scalebox{0.5}{\input{figures/GaussTwoLap.tex}}
    \caption{}
    \end{subfigure}
    \begin{subfigure}{0.33\linewidth}
    \scalebox{0.5}{\input{figures/GaussTwoGauss.tex}}
    \caption{}
    \end{subfigure}
    \caption{For the original data $X$ and $X'$ in (a) that is normal distributed with different mean and variance, (b) shows the resulting probability density of $Y = X + N$ and $Y' = X' + N$ for Laplace noise $N \sim \Lap(4)$, where the maximum logarithmic difference in probability density is $\max_{y} \left| \log \frac{P_{Y}(y)}{P_{Y'}(y)} \right| = 0.2992$. (c) shows the resulting probability density of $Y$ and $Y'$ for Gaussian noise $N \sim \Gauss(0,8)$, where $\max_{y} \left| \log \frac{P_{Y}(y)}{P_{Y'}(y)} \right| = 0.0156$. Note, the Laplace noise in (b) and Gaussian noise in (c) have the same variance.}
    \label{fig:Sample}
\end{figure*}

\section{PRELIMINARIES}
\label{sec:Prelim}

Denote $S$ the secret and $\SA$ the alphabet of the secret, where each $s \in \SA$ denotes an elementary event or outcome of $S$, e.g., $s = $``the patient has type 2 diabetes",  $s = $``the user is female", etc.
There is a statistical correlation between the data $X$ to be published and secret $S$, which can be described by the conditional probability $P_{X|S}(\cdot|s,\rho)$.
Here, $\rho$ denotes an adversary's prior belief on the probability distribution of $X$ given secret $S=s$, which could be different from others, i.e., for two adversaries obtaining prior beliefs $\rho$ and $\rho'$, $P_{X|S}(\cdot|s,\rho)$ and  $P_{X|S}(\cdot|s,\rho')$ are two different probability distributions even for the same secret $s$.
In this paper, $P_{X|S}(\cdot|s,\rho)$ is assumed to be Gaussian probability density function (See Section~\ref{sec:Monge}) characterized by its mean and variance. Thus, each adversary obtains his own prior belief $\rho$ constituted by pairs of means and variances, each of which corresponds to one secret $s$.

To protect secret $S$, a privatized version $Y$ of the original data $X$ is generated to ensure the indistinguishability between two distinct secrets $s_i$ and $s_j$.
Let $\SetPair \subseteq \SA^2$ be the \emph{discriminative pair set} containing all secret pairs $(s_i,s_j)$ on which a certain degree of statistical indistinguishability should be guaranteed against each adversary's prior belief $\rho$.
%

\begin{definition}[$(\epsilon,\delta)$-pufferfish privacy] \label{def}
	For $\delta \in [0,1)$ and $\epsilon > 0 $, $Y$ attains $\delta$-approximation of $\epsilon$-pufferfish privacy on the discriminative secret pair set $\SetPair$
	if for all $(s_i,s_j) \in \SetPair$ and $\rho$,
	\begin{equation} \label{eq:def_pufferfish}
		P_{Y|S}(B |s_i, \rho ) \leq e^{\epsilon}P_{Y|S}(B |s_j, \rho) + \delta, \ \forall B \subseteq \Real.
	\end{equation}
\end{definition}
If the condition~\eqref{eq:def_pufferfish} holds for $\delta = 0$, $(\epsilon, 0)$-pufferfish privacy is also called $\epsilon$-pufferfish privacy~\cite{Pufferfish2012KiferConf,Pufferfish2014Kifer}.

\subsection{Privatization mechanism}
We consider additive noise mechanism to generate $Y$.
Let $N$ be the zero mean noise that is independent of $X$. The goal is to attain pufferfish privacy in
$$ Y = X + N.$$
See Fig.~\ref{fig:Sample}.
Denoting $P_N(\cdot)$ the probability of $N$, we have the conditional probability
$$P_{Y|S}(y|s,\rho)= \int P_N(y-x) P_{X|S}(x|s,\rho) \dif x, $$
for which the  condition~\eqref{eq:def_pufferfish} is equivalent to
\begin{align}
	&P_{Y|S}(B|s_i, \rho )  - e^{\epsilon}P_{Y|S}(B|s_j, \rho) \nonumber \\
	& = \int_B \int P_{N}(y-x) P_{X|S}(x|s_i,\rho) \dif x \dif y\nonumber \\
	& \qquad\quad - e^{\epsilon} \int_B \int P_{N}(y-x') P_{X|S}(x'|s_j,\rho) \dif x' \dif y \nonumber \\
	& = \int_B \int \big( P_N(y-x) - e^\epsilon P_{N}(y-x') \big) \dif \pi(x,x') \dif y \leq \delta. \label{eq:mainInEqAux}
\end{align}
The coupling $\pi$ is the joint probability $\pi(x,x')$ with two marginals being $P_{X|S}(x|s_i,\rho)$ and $P_{X|S}(x'|s_j,\rho)$, i.e., $\int \pi(x,x') \dif x' = P_{X|S}(x|s_i,\rho)$ for all $x$ and $\int \pi(x,x') \dif x = P_{X|S}(x'|s_j,\rho)$ for all $x'$.

\subsection{Monge's Optimal Transport Plan $\hat{\pi}$}
\label{sec:Monge}
Assume that the priors are Gaussian distributed: $X|s_i \sim \Gauss(\mu_i, \sigma_i)$ and $X|s_j \sim \Gauss(\mu_j,\sigma_j)$ for each discriminative pair $(s_i,s_j) \in \SetPair$. Consider the Monge's optimal transport plan $\hat{\pi}$~\cite{Dowson1982Frechet,Givens1984NormClass,Takatsu2011Wnorm}:
$$\dif \hat{\pi}(x,x') = \dif P_{X|S}(x|s_i,\rho) \cdot \mathbb{I}\Set{x' = T(x)}$$
    where $\mathbb{I}\Set{\cdot}$ is the indicator function and $T$ is the linear mapping
    \begin{equation} \label{eq:Tmap}
        T(x) = \mu_j + \frac{\sigma_j}{\sigma_i} (x - \mu_i),\quad  \forall x.
    \end{equation}
Inequality \eqref{eq:mainInEqAux} under $\hat{\pi}$ reduces to
\begin{equation}
     \int \int_B \big( P_N(y-x) - e^\epsilon P_{N}(y-T(x)) \big) \dif y \dif P_{X|S}(x|s_i,\rho) \leq \delta.  \label{eq:mainInEq}
\end{equation}
The main results in this paper are derived by proving the inequality~\eqref{eq:mainInEq}. The transport plan $\hat{\pi}$ is in fact the minimizer of the Wasserstein metric $W_\alpha = \left( \inf_{\pi} \int d^{\alpha}(x,x') \dif \pi(x,x') \right)^{1/\alpha}$ in the order of $\alpha=2$, where the infimum is taken over all couplings $\pi$ of two Gaussian marginals~\cite{Takatsu2011Wnorm} \cite[Remark~2.31]{OTbook2017}.

\paragraph*{Noise reduction}
It is clear that  any coupling $\pi$ can be adopted to derive a sufficient condition for $(\epsilon,\delta)$-pufferfish privacy, e.g., determining the minimum noise power that satisfies \eqref{eq:mainInEqAux} under transport plan $\pi(x,x') = P_{X|S}(x|s_i,\rho)  P_{X|S}(x'|s_j,\rho), \forall x,x'$. But, it might result in a high noise power that jeopardizes data utility.
We adopt the $W_2$ or Monge's optimal transport method in an attempt to minimize the noise for attaining $(\epsilon,\delta)$-pufferfish privacy.
See Appendix~\ref{append:W2}, where we explain in detail how the minimization $\inf_{\pi}$ in Monge's optimal transport plan contributes to a noise reduction over all couplings.

\section{GAUSSIAN PRIORS}
\label{sec:LapGaussian}
%
We  consider Laplace noise $N \sim \Lap(b)$ with the noise distribution $P_{N}(z) = \frac{1}{2b} e^{-\frac{|z|}{b}}, \forall z \in \Real$.
\begin{theorem} \label{theo:Lap}
    For $X|s_i \sim \Gauss(\mu_i, \sigma_i)$ and $X|s_j \sim \Gauss(\mu_j,\sigma_j)$ for all $(s_i,s_j) \in \SetPair$, adding Laplace noise $N \sim \Lap (b) $ with
    \begin{equation}\label{eq:theo:Lap}
        b \geq \frac{1}{\epsilon} \max_{\rho,(s_i, s_j) \in \SetPair}  \big\{ |\mu_i - \mu_j| +|\sigma_i - \sigma_j|\tau^*(\delta) \big\}
    \end{equation}
	attains $(\epsilon, \delta)$-pufferfish private on $\SetPair$ in $Y$.
\end{theorem}

The proof is in Appendix~\ref{app:theo:Lap}.
In~\eqref{eq:theo:Lap}, $\tau^*(\delta) = \min\Set{\tau \colon \Pr(Z > \tau ) \leq \frac{\delta}{2}}$ or $\tau^*(\delta) = Q^{-1}(\frac{\delta}{2})$, where $Q(t)= \frac{1}{\sqrt{2\pi}} \int_{t}^{\infty} e^{-x^2/2} \dif x$ is the tail probability of standard normal distribution. Alternatively, one can derive the value of $\tau^*(\delta)$ by the Lambert-$W$ function. See Appendix~\ref{app:theo:Lap}.
%
Theorem~\ref{theo:Lap} states that to attain sufficient statistical indistinguishability in the released data, the additive noise should be large enough to compensate the difference in both mean and variance.
The second term $|\sigma_i - \sigma_j|\tau^*(\delta)$ in~\eqref{eq:theo:Lap} corresponds to the difference in variance.
The value of $\tau^*(\delta)$ is decreasing in $\delta \in (0,1)$ and can be obtained numerically. See Fig.~\ref{fig:TauStarDelta} in Appendix~\ref{app:theo:Lap}.

\subsection{Special Case: $\ell_1$-sensitivity Method for Differential Privacy}
\label{sec:Generalization}
We show below that Theorem~\ref{theo:Lap} is a generalization of the $\ell_1$-sensitivity noise calibration method proposed in~\cite{CalibNoiseDP} for attaining differential privacy (DP).
\begin{remark}[Translation priors] \label{coro:TransPrior}
    If for all $(s_i,s_j) \in \SetPair$, $X|s_i$ and $X|s_j$ are translation rvs to each other, i.e.,
    $P_{X|S}(x|s_i) = P_{X|S}(x - \mu_j + \mu_i | s_j)$ for all $x$, $Y = X + N$ for $N \sim \Lap (b)$ is $\epsilon$-pufferfish private if
    \begin{equation} \label{eq:Translation}
        b \geq \frac{1}{\epsilon} \max_{\rho,(s_i, s_j) \in \SetPair} |\mu_i - \mu_j|.
    \end{equation}
\end{remark}
Remark~\ref{coro:TransPrior} is derived by the fact that the second term $|\sigma_i - \sigma_j|\tau^*(\delta)$ in~\eqref{eq:theo:Lap} vanishes if $\sigma_i = \sigma_j$ for all $(s_i,s_j) \in \SetPair$.
The proof is in Appendix~\ref{app:coro:TransPrior} showing that Remark~\ref{coro:TransPrior} holds for any translation priors $X|s_i$ and $X|s_j$, not just Gaussian distribution.
A similar result can also be found in \cite[Theorem~4.1]{Chen2022AttrPriv}.

In DP~\cite{Dwork2014book}, data $X$ is deterministic, i.e., priors $P_{X|S}(.|s_i)$ and $P_{X|S}(.|s_j)$ are point masses located at means $\mu_i$ and $\mu_j$, respectively, with the same variance $\sigma_i=\sigma_j=0$, e.g., query answers of two neighboring databases.
In this case, the sufficient condition in Theorem~\ref{theo:Lap} reduces to
$ b \geq \frac{1}{\epsilon} \max_{\rho,(s_i, s_j) \in \SetPair}  |\mu_i - \mu_j|$, which is exactly the $\ell_1$-sensitivity method proposed in \cite{CalibNoiseDP} for attaining $\epsilon$-DP.
In other words, Theorem~\ref{theo:Lap} extends the $\ell_1$-sensitivity method for probabilistic priors with different variance.

\subsection{Summation query in $K$-independent user system}
\label{sec:CountQ}
Assume there are $K$ users indexed by $\K = \Set{ 1,\dotsc, K}$. Each user $k$ obtains a random variable $Z_k$, independently. We construct a multiple random variable $Z = (Z_k \colon k \in \K)$ for the overall outcome of the system.
The answer to the summation query is $X = \sum_{k \in \K } Z_k$.
Let the adversary's prior belief $\rho$ on $X$ be a Gaussian distribution with observed mean and variance.
Note, in this case, this prior belief is a good approximation of the true distribution of $X$ by the central limit theorem (CLT).

For each user $k$, denote ``$Z_k = \perp$" the event that user $k$ is absent in the system, ``$Z_k \neq \perp$" the event that user $k$ is present in the system and ``$Z_k = a$" the event that user is present in the system and reports the value $a$ of random variable $Z_k$.
Consider a discriminative pair set $\SetPair$ that consists of mutually exclusive secret pairs $s_i$ and $s_j$ denoting whether or not user $k$ is present, i.e., $s_i = ``Z_k = \perp"$ and $s_j=``Z_k \neq \perp"$. Alternatively, we define
$$ \SetPair_{\perp} = \big\{ (``Z_k \neq \perp", ``Z_k = \perp") \colon k \in \K \big\}. $$
It is clear that the pufferfish privacy on $\SetPair_{\perp}$ guarantees the adversary's indistinguishability between the existence and nonexistence of individual users.
Or, if the purpose is to make the actual realization of $Z_k$ indistinguishable for each user, we can define the discriminative pair set
$$ \SetPair_a = \big\{ (``Z_k = a", ``Z_k = a'") \colon k \in \K \big\} $$
for some $a, a' \in \Real$ such that $a \neq a'$.
For discrete rv $Z_k$, typically $a' = a \pm 1$.

In general, $Z_k$'s are mutually independent, but not necessarily identically distributed.
For each $k \in \K $, assume $Z_k$ is a random variable with mean $\mu_k$ and variance $\sigma_k^2$.
Let the adversary's prior belief $\rho$ be that the summation query
\begin{equation} \label{eq:KIndPrior}
    X = \sum_{k \in \K } Z_k \sim \Gauss \bigg( \sum_{k \in \K} \mu_k, \sum_{k \in \K} \sigma_k^2 \bigg)
\end{equation}
when all $K$ users present in the system.
Based on \eqref{eq:KIndPrior} and due to the independence of $Z_k$'s, given $Z_k = a$,
\begin{equation} \label{eq:KIndPriorAux}
    X = a + \sum_{k' \in \K_{-k}} Z_{k'} \sim \Gauss \left( a + \sum_{k' \in \K_{-k}} \mu_{k'}, \sum_{k' \in \K_{-k}} \sigma_{k'}^2 \right),
\end{equation}
where $\K_{-k} = \K \setminus \Set{k}$ containing all users except user $k$.
Note, the prior beliefs \eqref{eq:KIndPrior} and \eqref{eq:KIndPriorAux} approach the true probability distributions of $X$ for large $K$ based on CLT, if $Z_k$'s satisfy the Lyapunov or Lindeberg condition \cite{Fischer2011CLTbook}.

Using Theorem~\ref{theo:Lap}, we derive a sufficient condition for attaining pufferfish privacy on $\SetPair_{\perp}$ and $\SetPair_{a}$ below.

\begin{corollary} \label{coro:KIndLap}
In the $K$-independent user system, let $X$ be the summation query and $N \sim \Lap(b)$.
$Y = X + N$ is
\begin{enumerate}[(a)]
  \item $(\epsilon,\delta)$-pufferfish private on $\SetPair_{\perp}$ if
    \begin{equation}\label{eq:KIndLapA}
      b \geq \frac{1}{\epsilon} \max_{k \in \K} \big\{ |\mu_k| + \triangle \sigma_k \tau^*(\delta) \big\}
    \end{equation}
    where $\triangle\sigma_{k} = \sqrt{\sum_{k' \in \K_{-k}} \sigma_{k'}^2 + \sigma_k^2} - \sqrt{\sum_{k' \in \K_{-k}} \sigma_{k'}^2} $.
  \item $\epsilon$-pufferfish private on $\SetPair_a$  if
  \begin{equation}\label{eq:KIndLapB}
    b \geq \frac{|a-a'|}{\epsilon}.
  \end{equation}
\end{enumerate}
\end{corollary}
\begin{IEEEproof}
    For discriminative pair set $\SetPair_\perp$, given $Z_k = \perp$, $X$ has mean $\sum_{k' \in \K_{-k}} \mu_{k'}$ and variance $\sum_{k' \in \K_{-k}} \sigma_{k'}^2$;
    given $Z \neq \perp$, $X$ has mean $\sum_{k' \in \K_{-k}} \mu_{k'} + \mu_k$ and variance $\sum_{k' \in \K_{-k}} \sigma_{k'}^2 + \sigma_k^2$. Applying Theorem~\ref{theo:Lap}, we have (a).
    For discriminative pair set $\SetPair_a$ containing secret pairs $Z_k = a$ and $Z_k = a'$, we have $X$ given $Z_k = a$ being a translation of $X$ given $Z_k = a'$, i.e., $P_{X|S}(x|Z_k = a) = P_{X|S}(x-a'+a|Z_k = a'), \forall x \in \Real$, where the priors have the same variance but differ in mean. By Remark~\ref{coro:TransPrior}, calibrating noise to the difference in mean $|a-a'|$ attains $\epsilon$-pufferfish privacy.
\end{IEEEproof}

In Corollary~\ref{coro:KIndLap}(a), $\triangle\sigma_k \rightarrow 0 $ as $K \rightarrow \infty$.
That is, when the number of users $K$ grows,  it is getting more difficult for an adversary to distinguish the changes in variance of the summation query conditioned on the participation of individual users.
Equivalently, it is easier for an individual to hide his/her presence in a larger system. This coincides with the intuition of early definition of  data privacy, e.g.,  $K$-anonymity \cite{KAnony1998}, $t$-closeness \cite{TClose2007}, to guarantee a lower bound on the number of users where indistinguishability is achievable.
In addition, as $K \rightarrow \infty$, the lower bound in \eqref{eq:KIndLapA} approaches $\frac{1}{\epsilon}\max_{k \in \K} \mu_k$, i.e., it suffices to scale the Laplace noise to the maximum mean over all users.

\emph{Identical users}: If $Z_k$ is identically distributed, i.e., $\mu_k = \mu $ and $\sigma_k^2 = \sigma^2$ for all $k \in \K$, we have \eqref{eq:KIndLapA} reduced to
    \begin{equation} \label{eq:KIndLapC}
       b \geq \frac{1}{\epsilon} \Big( |\mu| + (\sqrt{K} - \sqrt{K-1}) \sigma \tau^*(\delta) \Big).
    \end{equation}
Here, the multiple random variable $Z = (Z_k \colon k \in \K)$ could also refer to a $K$-length \emph{i.i.d.} sample sequence, e.g., a dataset with $K$ records/rows. In this case, \eqref{eq:KIndLapC} states that as the number of records $K$ grows larger, the term $(\sqrt{K} - \sqrt{K-1}) \sigma \tau^*(\delta)$ vanishes, and we can attain $\epsilon$-pufferfish privacy by adding Laplace noise $N \sim \Lap(b)$ with $b = \mu / \epsilon$. See Fig.~\ref{fig:Lap_KIndSame}.
If $\mu \rightarrow 0$, only small amount of noise is required.
The identically distributed $Z_k$ setting also applies to data parallelism for distributed learning, e.g.,~\cite{Li2021}, where a large dataset is partitioned and assigned to users to allow parallel model training.

\begin{figure}[t]
\centerline{
\scalebox{0.6}{
%
%
\definecolor{crimson2143940}{RGB}{214,39,40}
\definecolor{darkgray176}{RGB}{176,176,176}
\definecolor{darkorange25512714}{RGB}{255,127,14}
\definecolor{forestgreen4416044}{RGB}{44,160,44}
\definecolor{mediumpurple148103189}{RGB}{148,103,189}
\definecolor{steelblue31119180}{RGB}{31,119,180}
\begin{tikzpicture}

\begin{axis}[%
width=4.3in,
height=2in,
scale only axis,
xmin=1,
xmax=50,
xlabel = {\Large the number of users $K$},
ymin=1,
ymax=5,
ylabel = {$\Big( \mu + (\sqrt{K} - \sqrt{K-1}) \sigma \tau^*(\delta) \Big) / \epsilon$},
grid=major,
legend style={at={(0.98,0.97)},draw=darkgray!60!black,fill=white,legend cell align=left}
]
\addplot [very thick, steelblue31119180]
table {%
1 2.23737633763376
2 1.51253806080745
3 1.39328428642396
4 1.33155399040235
5 1.2921049294313
6 1.26408304229769
7 1.24284942057831
8 1.22603872930761
9 1.21230021601885
10 1.20079853681893
11 1.1909854866751
12 1.18248433333505
13 1.17502646265402
14 1.16841448154228
15 1.16249963472003
16 1.15716740188836
17 1.15232798816889
18 1.14790985661971
19 1.14385521555039
20 1.14011679852361
21 1.1366555209563
22 1.13343874451605
23 1.13043897135524
24 1.12763284776779
25 1.12500039417578
26 1.12252440308372
27 1.12018996334301
28 1.11798408055412
29 1.11589537145666
30 1.11391381584508
31 1.11203055363499
32 1.11023771767848
33 1.10852829511607
34 1.10689601168139
35 1.10533523459862
36 1.10384089064163
37 1.10240839663371
38 1.10103360021574
39 1.09971272913635
40 1.09844234765204
41 1.09721931888821
42 1.09604077222135
43 1.09490407490936
44 1.09380680733127
45 1.09274674130585
46 1.09172182104673
47 1.0907301463831
48 1.08976995793441
49 1.08883962397563
50 1.08793762876975
};
\addlegendentry{\large $\mu = 1, \sigma^2=1$};

\addplot [very thick, darkorange25512714]
table {%
1 3.47475267526753
2 2.02507612161491
3 1.78656857284792
4 1.66310798080469
5 1.5842098588626
6 1.52816608459538
7 1.48569884115663
8 1.45207745861521
9 1.42460043203771
10 1.40159707363785
11 1.38197097335019
12 1.36496866667009
13 1.35005292530803
14 1.33682896308456
15 1.32499926944007
16 1.31433480377672
17 1.30465597633779
18 1.29581971323941
19 1.28771043110077
20 1.28023359704723
21 1.27331104191261
22 1.26687748903209
23 1.26087794271049
24 1.25526569553558
25 1.25000078835156
26 1.24504880616743
27 1.24037992668601
28 1.23596816110825
29 1.23179074291332
30 1.22782763169017
31 1.22406110726998
32 1.22047543535695
33 1.21705659023214
34 1.21379202336277
35 1.21067046919724
36 1.20768178128326
37 1.20481679326743
38 1.20206720043149
39 1.19942545827271
40 1.19688469530408
41 1.19443863777642
42 1.19208154444271
43 1.18980814981873
44 1.18761361466253
45 1.18549348261171
46 1.18344364209346
47 1.18146029276619
48 1.17953991586883
49 1.17767924795125
50 1.17587525753949
};
\addlegendentry{\large $\mu = 1, \sigma^2=4$};

\addplot [very thick, forestgreen4416044]
table {%
1 4.71212901290129
2 2.53761418242236
3 2.17985285927188
4 1.99466197120704
5 1.8763147882939
6 1.79224912689307
7 1.72854826173494
8 1.67811618792282
9 1.63690064805656
10 1.60239561045678
11 1.57295646002529
12 1.54745300000514
13 1.52507938796205
14 1.50524344462685
15 1.4874989041601
16 1.47150220566509
17 1.45698396450668
18 1.44372956985912
19 1.43156564665116
20 1.42035039557084
21 1.40996656286891
22 1.40031623354814
23 1.39131691406573
24 1.38289854330337
25 1.37500118252734
26 1.36757320925115
27 1.36056989002902
28 1.35395224166237
29 1.34768611436998
30 1.34174144753525
31 1.33609166090496
32 1.33071315303543
33 1.32558488534822
34 1.32068803504416
35 1.31600570379586
36 1.31152267192488
37 1.30722518990114
38 1.30310080064723
39 1.29913818740906
40 1.29532704295612
41 1.29165795666463
42 1.28812231666406
43 1.28471222472809
44 1.2814204219938
45 1.27824022391756
46 1.27516546314019
47 1.27219043914929
48 1.26930987380324
49 1.26651887192688
50 1.26381288630924
};
\addlegendentry{\large $\mu = 1, \sigma^2=9$};

\addplot [very thick, crimson2143940]
table {%
1 5.94950535053505
2 3.05015224322982
3 2.57313714569584
4 2.32621596160939
5 2.1684197177252
6 2.05633216919077
7 1.97139768231325
8 1.90415491723042
9 1.84920086407541
10 1.8031941472757
11 1.76394194670039
12 1.72993733334018
13 1.70010585061606
14 1.67365792616913
15 1.64999853888014
16 1.62866960755345
17 1.60931195267558
18 1.59163942647883
19 1.57542086220154
20 1.56046719409445
21 1.54662208382521
22 1.53375497806418
23 1.52175588542098
24 1.51053139107116
25 1.50000157670313
26 1.49009761233487
27 1.48075985337202
28 1.47193632221649
29 1.46358148582665
30 1.45565526338034
31 1.44812221453995
32 1.44095087071391
33 1.43411318046429
34 1.42758404672554
35 1.42134093839448
36 1.41536356256651
37 1.40963358653486
38 1.40413440086298
39 1.39885091654542
40 1.39376939060816
41 1.38887727555284
42 1.38416308888542
43 1.37961629963745
44 1.37522722932507
45 1.37098696522341
46 1.36688728418692
47 1.36292058553239
48 1.35907983173765
49 1.3553584959025
50 1.35175051507899
};
\addlegendentry{\large $\mu = 1, \sigma^2=16$};

\addplot [very thick, mediumpurple148103189]
table {%
1 7.18688168816882
2 3.56269030403727
3 2.96642143211981
4 2.65776995201174
5 2.4605246471565
6 2.32041521148846
7 2.21424710289157
8 2.13019364653803
9 2.06150108009427
10 2.00399268409463
11 1.95492743337548
12 1.91242166667523
13 1.87513231327008
14 1.84207240771141
15 1.81249817360017
16 1.78583700944181
17 1.76163994084447
18 1.73954928309853
19 1.71927607775193
20 1.70058399261807
21 1.68327760478151
22 1.66719372258023
23 1.65219485677622
24 1.63816423883895
25 1.62500197087891
26 1.61262201541858
27 1.60094981671503
28 1.58992040277062
29 1.57947685728331
30 1.56956907922542
31 1.56015276817494
32 1.55118858839238
33 1.54264147558036
34 1.53448005840693
35 1.52667617299311
36 1.51920445320814
37 1.51204198316857
38 1.50516800107872
39 1.49856364568177
40 1.4922117382602
41 1.48609659444105
42 1.48020386110677
43 1.47452037454681
44 1.46903403665633
45 1.46373370652927
46 1.45860910523365
47 1.45365073191548
48 1.44884978967206
49 1.44419811987813
50 1.43968814384874
};
\addlegendentry{\large $\mu = 1, \sigma^2=25$};

\end{axis}
\end{tikzpicture}%
}
}
\caption{For the $K$-independent and identical user system, the lower bound on $b$ in \eqref{eq:KIndLapC} for attaining $(1, 0.3)$-pufferfish privacy with Laplace noise $N \sim \Lap(b)$ as the number of users $K$ increases. We set the mean $\mu = 1$ and vary the variance $\sigma^2$ from $1$ to $25$.} \label{fig:Lap_KIndSame}
\end{figure}
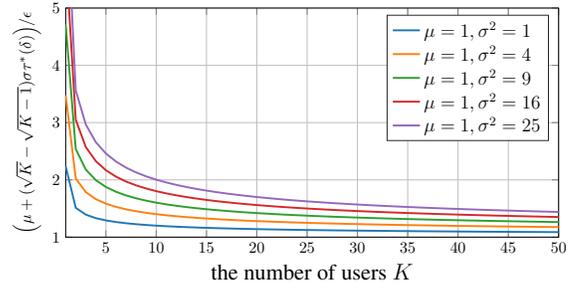

\begin{figure*}[h]
	\begin{subfigure}{0.33\linewidth}
		\scalebox{0.77}{
\begin{tikzpicture}

\definecolor{darkgray176}{RGB}{176,176,176}
\definecolor{darkorange25512714}{RGB}{255,127,14}
\definecolor{steelblue31119180}{RGB}{151,160,190}

\begin{axis}[
width=3in,
height=2.3in,
tick align=outside,
tick pos=left,
x grid style={darkgray176},
xmin=-2, xmax=22,
xtick style={color=black},
y grid style={darkgray176},
ymin=0, ymax=0.24,
ytick style={color=black},
yticklabel style={
        /pgf/number format/fixed,
        /pgf/number format/precision=3
},
ytick={0.05,0.1,0.15,0.2},
legend style={at={(0.98,0.97)},draw=darkgray!60!black,fill=white,legend cell align=left,nodes={scale=0.7, transform shape}},
]
\draw[draw=none,fill=steelblue31119180] (axis cs:-0.677489006874814,0) rectangle (axis cs:0.295731453065696,0.00131564201927303);
\draw[draw=none,fill=steelblue31119180] (axis cs:0.295731453065696,0) rectangle (axis cs:1.26895191300621,0.00164455252409128);
\draw[draw=none,fill=steelblue31119180] (axis cs:1.26895191300621,0) rectangle (axis cs:2.24217237294672,0.00361801555300083);
\draw[draw=none,fill=steelblue31119180] (axis cs:2.24217237294672,0) rectangle (axis cs:3.21539283288723,0.010196225649366);
\draw[draw=none,fill=steelblue31119180] (axis cs:3.21539283288723,0) rectangle (axis cs:4.18861329282773,0.0118407781734573);
\draw[draw=none,fill=steelblue31119180] (axis cs:4.18861329282773,0) rectangle (axis cs:5.16183375276824,0.0223659143276415);
\draw[draw=none,fill=steelblue31119180] (axis cs:5.16183375276824,0) rectangle (axis cs:6.13505421270875,0.0414427236071004);
\draw[draw=none,fill=steelblue31119180] (axis cs:6.13505421270875,0) rectangle (axis cs:7.10827467264926,0.0555858753142854);
\draw[draw=none,fill=steelblue31119180] (axis cs:7.10827467264926,0) rectangle (axis cs:8.08149513258977,0.0920949413491119);
\draw[draw=none,fill=steelblue31119180] (axis cs:8.08149513258977,0) rectangle (axis cs:9.05471559253028,0.185834435222315);
\draw[draw=none,fill=steelblue31119180] (axis cs:9.05471559253028,0) rectangle (axis cs:10.0279360524708,0.216423112170413);
\draw[draw=none,fill=steelblue31119180] (axis cs:10.0279360524708,0) rectangle (axis cs:11.0011565124113,0.151627742721216);
\draw[draw=none,fill=steelblue31119180] (axis cs:11.0011565124113,0) rectangle (axis cs:11.9743769723518,0.0733470425744712);
\draw[draw=none,fill=steelblue31119180] (axis cs:11.9743769723518,0) rectangle (axis cs:12.9475974322923,0.0651242799540149);
\draw[draw=none,fill=steelblue31119180] (axis cs:12.9475974322923,0) rectangle (axis cs:13.9208178922328,0.0522967702661028);
\draw[draw=none,fill=steelblue31119180] (axis cs:13.9208178922328,0) rectangle (axis cs:14.8940383521733,0.0263128403854605);
\draw[draw=none,fill=steelblue31119180] (axis cs:14.8940383521733,0) rectangle (axis cs:15.8672588121139,0.0134853306975485);
\draw[draw=none,fill=steelblue31119180] (axis cs:15.8672588121139,0) rectangle (axis cs:16.8404792720544,0.00131564201927303);
\draw[draw=none,fill=steelblue31119180] (axis cs:16.8404792720544,0) rectangle (axis cs:17.8136997319949,0.00131564201927303);
\draw[draw=none,fill=steelblue31119180] (axis cs:17.8136997319949,0) rectangle (axis cs:18.7869201919354,0.000328910504818257);
\addlegendimage{area legend,fill=steelblue31119180,draw = white}
\addlegendentry{original data}

\addplot [very thick, blue]
table {%
0 0.000335674490852774
0.1 0.000393047639378568
0.2 0.000458939641363292
0.3 0.000534379116777324
0.4 0.000620478746402267
0.5 0.00071843563960535
0.6 0.000829530489037911
0.7 0.000955125332374754
0.8 0.00109665974190104
0.9 0.00125564526785878
1 0.00143365797153672
1.1 0.00163232889961148
1.2 0.00185333237262211
1.3 0.00209837198796428
1.4 0.00236916427156362
1.5 0.00266741995240092
1.6 0.0029948228800971
1.7 0.00335300665739612
1.8 0.00374352911596234
1.9 0.00416784482455603
2 0.00462727588226375
2.1 0.00512298131471007
2.2 0.00565592545654089
2.3 0.00622684576724752
2.4 0.0068362205877726
2.5 0.00748423740041073
2.6 0.00817076220239909
2.7 0.00889531064249731
2.8 0.00965702159820654
2.9 0.0104546338878696
3 0.0112864668160727
3.1 0.0121504052426726
3.2 0.0130438898466281
3.3 0.0139639132282942
3.4 0.014907022462521
3.5 0.0158693286867395
3.6 0.0168465242931288
3.7 0.0178339083052902
3.8 0.0188264205747759
3.9 0.0198186855524071
4 0.0208050665979573
4.1 0.0217797321158117
4.2 0.0227367352690885
4.3 0.0236701096504395
4.4 0.0245739840821634
4.5 0.0254427206678417
4.6 0.0262710812760527
4.7 0.0270544287114707
4.8 0.0277889697680709
4.9 0.0284720479409826
5 0.0291024934994007
5.1 0.0296810375229792
5.2 0.0302107939563191
5.3 0.0306978093058878
5.4 0.0311516729062604
5.5 0.0315861714699216
5.6 0.0320199599027478
5.7 0.0324772064732221
5.8 0.0329881551993825
5.9 0.0335895331540118
6 0.0343247172623429
6.1 0.0352435665813499
6.2 0.0364018248643519
6.3 0.0378600073418904
6.4 0.0396817076278402
6.5 0.0419312971225338
6.6 0.0446710403866534
6.7 0.0479577138377386
6.8 0.0518388875022463
6.9 0.0563491036269804
7 0.0615062526033241
7.1 0.0673084951696497
7.2 0.0737320990740854
7.3 0.0807305382824298
7.4 0.088235136370208
7.5 0.0961574207681768
7.6 0.104393195308503
7.7 0.112828146716167
7.8 0.121344595366637
7.9 0.129828807010917
8 0.138178129117331
8.1 0.146307131797003
8.2 0.154151943189369
8.3 0.161672087605119
8.4 0.168849363123833
8.5 0.17568361901249
8.6 0.18218568085351
8.7 0.188368076713655
8.8 0.194234585612249
8.9 0.199769902460279
9 0.20493084090879
9.1 0.209640442435311
9.2 0.213786115206526
9.3 0.217222506348007
9.4 0.21977926083204
9.5 0.221273207297627
9.6 0.22152391817394
9.7 0.220371103105349
9.8 0.217691984262328
9.9 0.21341671958072
10 0.20754010293527
10.1 0.200128160460611
10.2 0.191318827908479
10.3 0.181316557837078
10.4 0.170381377177923
10.5 0.15881350610739
10.6 0.146935083895218
10.7 0.135070777943656
10.8 0.123529061696051
10.9 0.112585750729244
11 0.102471026678029
11.1 0.0933607166329684
11.2 0.0853720997400818
11.3 0.0785640478544656
11.4 0.0729409256132956
11.5 0.0684594110513387
11.6 0.065037264077209
11.7 0.0625630606872888
11.8 0.0609060045251557
11.9 0.0599250937341655
12 0.0594771261672236
12.1 0.0594232382006004
12.2 0.0596338660056953
12.3 0.0599921757606534
12.4 0.0603961222141558
12.5 0.0607593622669492
12.6 0.0610112767260689
12.7 0.0610963476922465
12.8 0.0609731112524996
12.9 0.0606128650951741
13 0.0599982666870946
13.1 0.0591219159634311
13.2 0.0579849810121599
13.3 0.0565958978567974
13.4 0.0549691563963966
13.5 0.0531241729811895
13.6 0.0510842444964581
13.7 0.0488755774933064
13.8 0.0465263872183518
13.9 0.0440660639858087
14 0.0415244071711048
14.1 0.0389309294941315
14.2 0.0363142358163959
14.3 0.0337014812572645
14.4 0.0311179130738333
14.5 0.0285864995936874
14.6 0.0261276477476872
14.7 0.0237590086493456
14.8 0.0214953684290383
14.9 0.0193486193499797
15 0.0173278042671268
15.1 0.0154392258578887
15.2 0.0136866108318333
15.3 0.0120713185544259
15.4 0.0105925832023379
15.5 0.00924777868189449
15.6 0.00803269604245171
15.7 0.00694182394153477
15.8 0.00596862379710528
15.9 0.00510579251857655
16 0.00434550706690444
16.1 0.00367964648459404
16.2 0.00309998839594738
16.3 0.00259837825359394
16.4 0.00216687075798881
16.5 0.0017978438727383
16.6 0.00148408668258297
16.7 0.00121886298564845
16.8 0.000995952979601977
16.9 0.000809675702814601
17 0.000654895042655442
17.1 0.000527012143873744
17.2 0.000421946963257384
17.3 0.000336111545796363
17.4 0.00026637736530951
17.5 0.000210038800225675
17.6 0.000164774521931027
17.7 0.000128608274965382
17.8 9.98702384804011e-05
17.9 7.71598867870013e-05
18 5.93110206096674e-05
18.1 4.53594242629271e-05
18.2 3.45134194986446e-05
18.3 2.61274344719382e-05
18.4 1.96785848844109e-05
18.5 1.4746171548176e-05
18.6 1.09939313173688e-05
18.7 8.15483309999323e-06
18.8 6.01818390060659e-06
18.9 4.41879803286409e-06
19 3.22798248298419e-06
19.1 2.34609992938847e-06
19.2 1.69648554419399e-06
19.3 1.22051224254425e-06
19.4 8.73619735581302e-07
19.5 6.22144194005792e-07
19.6 4.40806498435254e-07
19.7 3.10737196924491e-07
19.8 2.17934917716922e-07
19.9 1.52071809472946e-07
};
\addlegendentry{GMM fitting};
\end{axis}

\end{tikzpicture}}
		\caption{$X|s_i$ and GMM fitting}
	\end{subfigure}
	\begin{subfigure}{0.33\linewidth}
		\scalebox{0.77}{
\begin{tikzpicture}

\definecolor{darkgray176}{RGB}{176,176,176}
\definecolor{steelblue31119180}{RGB}{151,160,220}

\begin{axis}[
width=3in,
height=2.3in,
tick align=outside,
tick pos=left,
x grid style={darkgray176},
xmin=-2, xmax=22,
xtick style={color=black},
y grid style={darkgray176},
ymin=0, ymax=0.19,
ytick style={color=black},
yticklabel style={
        /pgf/number format/fixed,
        /pgf/number format/precision=3
},
legend style={at={(0.98,0.97)},draw=darkgray!60!black,fill=white,legend cell align=left,nodes={scale=0.7, transform shape}},
]
\draw[draw=none,fill=steelblue31119180] (axis cs:-5,0) rectangle (axis cs:-4,0.000320102432778489);
\draw[draw=none,fill=steelblue31119180] (axis cs:-4,0) rectangle (axis cs:-3,0.000320102432778489);
\draw[draw=none,fill=steelblue31119180] (axis cs:-3,0) rectangle (axis cs:-2,0.00192061459667093);
\draw[draw=none,fill=steelblue31119180] (axis cs:-2,0) rectangle (axis cs:-1,0.00160051216389245);
\draw[draw=none,fill=steelblue31119180] (axis cs:-1,0) rectangle (axis cs:0,0.00192061459667093);
\draw[draw=none,fill=steelblue31119180] (axis cs:0,0) rectangle (axis cs:1,0.00416133162612036);
\draw[draw=none,fill=steelblue31119180] (axis cs:1,0) rectangle (axis cs:2,0.00448143405889885);
\draw[draw=none,fill=steelblue31119180] (axis cs:2,0) rectangle (axis cs:3,0.0121638924455826);
\draw[draw=none,fill=steelblue31119180] (axis cs:3,0) rectangle (axis cs:4,0.0169654289372599);
\draw[draw=none,fill=steelblue31119180] (axis cs:4,0) rectangle (axis cs:5,0.030089628681178);
\draw[draw=none,fill=steelblue31119180] (axis cs:5,0) rectangle (axis cs:6,0.0412932138284251);
\draw[draw=none,fill=steelblue31119180] (axis cs:6,0) rectangle (axis cs:7,0.0653008962868118);
\draw[draw=none,fill=steelblue31119180] (axis cs:7,0) rectangle (axis cs:8,0.0950704225352113);
\draw[draw=none,fill=steelblue31119180] (axis cs:8,0) rectangle (axis cs:9,0.149167733674776);
\draw[draw=none,fill=steelblue31119180] (axis cs:9,0) rectangle (axis cs:10,0.167733674775928);
\draw[draw=none,fill=steelblue31119180] (axis cs:10,0) rectangle (axis cs:11,0.12708066581306);
\draw[draw=none,fill=steelblue31119180] (axis cs:11,0) rectangle (axis cs:12,0.0829065300896287);
\draw[draw=none,fill=steelblue31119180] (axis cs:12,0) rectangle (axis cs:13,0.074583866837388);
\draw[draw=none,fill=steelblue31119180] (axis cs:13,0) rectangle (axis cs:14,0.0528169014084507);
\draw[draw=none,fill=steelblue31119180] (axis cs:14,0) rectangle (axis cs:15,0.0355313700384123);
\draw[draw=none,fill=steelblue31119180] (axis cs:15,0) rectangle (axis cs:16,0.015044814340589);
\draw[draw=none,fill=steelblue31119180] (axis cs:16,0) rectangle (axis cs:17,0.0089628681177977);
\draw[draw=none,fill=steelblue31119180] (axis cs:17,0) rectangle (axis cs:18,0.00512163892445583);
\draw[draw=none,fill=steelblue31119180] (axis cs:18,0) rectangle (axis cs:19,0.0028809218950064);
\draw[draw=none,fill=steelblue31119180] (axis cs:19,0) rectangle (axis cs:20,0.000960307298335467);
\draw[draw=none,fill=steelblue31119180] (axis cs:20,0) rectangle (axis cs:21,0.000960307298335467);
\draw[draw=none,fill=steelblue31119180] (axis cs:21,0) rectangle (axis cs:22,0.000640204865556978);
\draw[draw=none,fill=steelblue31119180] (axis cs:22,0) rectangle (axis cs:23,0);
\draw[draw=none,fill=steelblue31119180] (axis cs:23,0) rectangle (axis cs:24,0);
\draw[draw=none,fill=steelblue31119180] (axis cs:24,0) rectangle (axis cs:25,0);
\addlegendimage{area legend,fill=steelblue31119180,draw = white}
\addlegendentry[align=left]{noised data \\ $\epsilon=1$, $\delta=0.5$}
\end{axis}

\end{tikzpicture}}
		\caption{$Y|s_i$: $\epsilon = 1$ and $\delta=0.5$}
	\end{subfigure}
		\begin{subfigure}{0.33\linewidth}
		\scalebox{0.77}{
\begin{tikzpicture}

\definecolor{darkgray176}{RGB}{176,176,176}
\definecolor{steelblue31119180}{RGB}{151,160,220}

\begin{axis}[
width=3in,
height=2.3in,
tick align=outside,
tick pos=left,
x grid style={darkgray176},
xmin=-2, xmax=22,
xtick style={color=black},
y grid style={darkgray176},
ymin=0, ymax=0.140516688061617,
ytick style={color=black},
yticklabel style={
        /pgf/number format/fixed,
        /pgf/number format/precision=3
},
legend style={at={(0.98,0.97)},draw=darkgray!60!black,fill=white,legend cell align=left,nodes={scale=0.7, transform shape}},
]
\draw[draw=none,fill=steelblue31119180] (axis cs:-5,0) rectangle (axis cs:-4,0.000962772785622593);
\draw[draw=none,fill=steelblue31119180] (axis cs:-4,0) rectangle (axis cs:-3,0.00128369704749679);
\draw[draw=none,fill=steelblue31119180] (axis cs:-3,0) rectangle (axis cs:-2,0.00160462130937099);
\draw[draw=none,fill=steelblue31119180] (axis cs:-2,0) rectangle (axis cs:-1,0.00353016688061617);
\draw[draw=none,fill=steelblue31119180] (axis cs:-1,0) rectangle (axis cs:0,0.00385109114249037);
\draw[draw=none,fill=steelblue31119180] (axis cs:0,0) rectangle (axis cs:1,0.0115532734274711);
\draw[draw=none,fill=steelblue31119180] (axis cs:1,0) rectangle (axis cs:2,0.00930680359435173);
\draw[draw=none,fill=steelblue31119180] (axis cs:2,0) rectangle (axis cs:3,0.0163671373555841);
\draw[draw=none,fill=steelblue31119180] (axis cs:3,0) rectangle (axis cs:4,0.0279204107830552);
\draw[draw=none,fill=steelblue31119180] (axis cs:4,0) rectangle (axis cs:5,0.0346598202824134);
\draw[draw=none,fill=steelblue31119180] (axis cs:5,0) rectangle (axis cs:6,0.0452503209242619);
\draw[draw=none,fill=steelblue31119180] (axis cs:6,0) rectangle (axis cs:7,0.0754172015404365);
\draw[draw=none,fill=steelblue31119180] (axis cs:7,0) rectangle (axis cs:8,0.0959563543003851);
\draw[draw=none,fill=steelblue31119180] (axis cs:8,0) rectangle (axis cs:9,0.109756097560976);
\draw[draw=none,fill=steelblue31119180] (axis cs:9,0) rectangle (axis cs:10,0.13382541720154);
\draw[draw=none,fill=steelblue31119180] (axis cs:10,0) rectangle (axis cs:11,0.10783055198973);
\draw[draw=none,fill=steelblue31119180] (axis cs:11,0) rectangle (axis cs:12,0.0872913992297818);
\draw[draw=none,fill=steelblue31119180] (axis cs:12,0) rectangle (axis cs:13,0.061617458279846);
\draw[draw=none,fill=steelblue31119180] (axis cs:13,0) rectangle (axis cs:14,0.0564826700898588);
\draw[draw=none,fill=steelblue31119180] (axis cs:14,0) rectangle (axis cs:15,0.0391527599486521);
\draw[draw=none,fill=steelblue31119180] (axis cs:15,0) rectangle (axis cs:16,0.02599486521181);
\draw[draw=none,fill=steelblue31119180] (axis cs:16,0) rectangle (axis cs:17,0.0163671373555841);
\draw[draw=none,fill=steelblue31119180] (axis cs:17,0) rectangle (axis cs:18,0.0112323491655969);
\draw[draw=none,fill=steelblue31119180] (axis cs:18,0) rectangle (axis cs:19,0.0112323491655969);
\draw[draw=none,fill=steelblue31119180] (axis cs:19,0) rectangle (axis cs:20,0.00417201540436457);
\draw[draw=none,fill=steelblue31119180] (axis cs:20,0) rectangle (axis cs:21,0.00288831835686778);
\draw[draw=none,fill=steelblue31119180] (axis cs:21,0) rectangle (axis cs:22,0.00128369704749679);
\draw[draw=none,fill=steelblue31119180] (axis cs:22,0) rectangle (axis cs:23,0.00192554557124519);
\draw[draw=none,fill=steelblue31119180] (axis cs:23,0) rectangle (axis cs:24,0.000962772785622593);
\draw[draw=none,fill=steelblue31119180] (axis cs:24,0) rectangle (axis cs:25,0.000320924261874198);
\addlegendimage{area legend,fill=steelblue31119180,draw = white}
\addlegendentry[align=left]{noised data \\ $\epsilon=1$, $\delta=0.3$}
\end{axis}

\end{tikzpicture}}
		\caption{$Y|s_i$: $\epsilon = 1$ and $\delta=0.3$}
	\end{subfigure} \\
	\begin{subfigure}{0.33\linewidth}
		\scalebox{0.77}{
\begin{tikzpicture}

\definecolor{darkgray176}{RGB}{176,176,176}
\definecolor{darkorange25512714}{RGB}{190,16,16}
\definecolor{steelblue31119180}{RGB}{185,145,145}

\begin{axis}[
width=3in,
height=2.3in,
tick align=outside,
tick pos=left,
x grid style={darkgray176},
xmin=-2, xmax=22,
xtick style={color=black},
y grid style={darkgray176},
ymin=0, ymax=0.19,
ytick style={color=black},
ytick={0.05,0.1,0.15,0.2},
yticklabel style={
        /pgf/number format/fixed,
        /pgf/number format/precision=3
},
legend style={at={(0.4,0.97)},draw=darkgray!60!black,fill=white,legend cell align=left,nodes={scale=0.7, transform shape}}
]
\draw[draw=none,fill=steelblue31119180] (axis cs:0.618052535163751,0) rectangle (axis cs:1.49184173596531,0.0055074147558734);
\draw[draw=none,fill=steelblue31119180] (axis cs:1.49184173596531,0) rectangle (axis cs:2.36563093676687,0.00881186360939745);
\draw[draw=none,fill=steelblue31119180] (axis cs:2.36563093676687,0) rectangle (axis cs:3.23942013756842,0.0121163124629215);
\draw[draw=none,fill=steelblue31119180] (axis cs:3.23942013756842,0) rectangle (axis cs:4.11320933836998,0.0154207613164455);
\draw[draw=none,fill=steelblue31119180] (axis cs:4.11320933836998,0) rectangle (axis cs:4.98699853917154,0.00771038065822276);
\draw[draw=none,fill=steelblue31119180] (axis cs:4.98699853917154,0) rectangle (axis cs:5.86078773997309,0.0110148295117468);
\draw[draw=none,fill=steelblue31119180] (axis cs:5.86078773997309,0) rectangle (axis cs:6.73457694077465,0.0198266931211442);
\draw[draw=none,fill=steelblue31119180] (axis cs:6.73457694077465,0) rectangle (axis cs:7.60836614157621,0.0264355908281923);
\draw[draw=none,fill=steelblue31119180] (axis cs:7.60836614157621,0) rectangle (axis cs:8.48215534237777,0.079306772484577);
\draw[draw=none,fill=steelblue31119180] (axis cs:8.48215534237777,0) rectangle (axis cs:9.35594454317932,0.132177954140962);
\draw[draw=none,fill=steelblue31119180] (axis cs:9.35594454317932,0) rectangle (axis cs:10.2297337439809,0.158613544969154);
\draw[draw=none,fill=steelblue31119180] (axis cs:10.2297337439809,0) rectangle (axis cs:11.1035229447824,0.110148295117468);
\draw[draw=none,fill=steelblue31119180] (axis cs:11.1035229447824,0) rectangle (axis cs:11.977312145584,0.0947275338010225);
\draw[draw=none,fill=steelblue31119180] (axis cs:11.977312145584,0) rectangle (axis cs:12.8511013463856,0.111249778068643);
\draw[draw=none,fill=steelblue31119180] (axis cs:12.8511013463856,0) rectangle (axis cs:13.7248905471871,0.152004647262106);
\draw[draw=none,fill=steelblue31119180] (axis cs:13.7248905471871,0) rectangle (axis cs:14.5986797479887,0.0936260508498479);
\draw[draw=none,fill=steelblue31119180] (axis cs:14.5986797479887,0) rectangle (axis cs:15.4724689487902,0.0704949088751796);
\draw[draw=none,fill=steelblue31119180] (axis cs:15.4724689487902,0) rectangle (axis cs:16.3462581495918,0.0198266931211443);
\draw[draw=none,fill=steelblue31119180] (axis cs:16.3462581495918,0) rectangle (axis cs:17.2200473503933,0.00881186360939741);
\draw[draw=none,fill=steelblue31119180] (axis cs:17.2200473503933,0) rectangle (axis cs:18.0938365511949,0.00660889770704808);
\addlegendimage{area legend,fill=steelblue31119180,draw = white}
\addlegendentry{original data}

\addplot [very thick, darkorange25512714]
table {%
0 0.00136404973296059
0.1 0.00151944122384611
0.2 0.00168827664343639
0.3 0.00187115311309589
0.4 0.00206862156492531
0.5 0.00228117590745662
0.6 0.00250924178833351
0.7 0.00275316508965846
0.8 0.00301320031073452
0.9 0.00328949901066817
1 0.00358209849917357
1.1 0.00389091097737167
1.2 0.00421571334085755
1.3 0.00455613786429554
1.4 0.00491166398981128
1.5 0.0052816114400679
1.6 0.00566513487079792
1.7 0.0060612202664726
1.8 0.00646868326659111
1.9 0.00688616958875585
2 0.00731215768839091
2.1 0.00774496376392508
2.2 0.00818274918090993
2.3 0.00862353034943518
2.4 0.00906519104703959
2.5 0.00950549713493609
2.6 0.00994211356975028
2.7 0.0103726235672106
2.8 0.0107945497295522
2.9 0.011205376906151
3 0.01160257651857
3.1 0.0119836320484058
3.2 0.0123460653608882
3.3 0.0126874635211694
3.4 0.0130055057559972
3.5 0.0132979902237931
3.6 0.0135628602843116
3.7 0.0137982300089706
3.8 0.0140024087492259
3.9 0.014173924688427
4 0.014311547448557
4.1 0.0144143100137887
4.2 0.0144815304746794
4.3 0.0145128343962434
4.4 0.0145081789745259
4.5 0.0144678805707285
4.6 0.014392647694957
4.7 0.0142836220406176
4.8 0.0141424307213452
4.9 0.013971253396306
5 0.0137729084299544
5.1 0.0135509625413418
5.2 0.0133098684565305
5.3 0.0130551347653821
5.4 0.0127935313642907
5.5 0.0125333323946765
5.6 0.0122845963241372
5.7 0.0120594796483945
5.8 0.0118725765512651
5.9 0.0117412717568024
6 0.0116860878582923
6.1 0.0117310018624537
6.2 0.0119036989488341
6.3 0.012235725084296
6.4 0.0127624948793548
6.5 0.0135231077847373
6.6 0.014559925334519
6.7 0.0159178655697208
6.8 0.0176433788309135
6.9 0.0197830823572294
7 0.0223820497667502
7.1 0.025481775218387
7.2 0.029117859975198
7.3 0.0333174996697829
7.4 0.0380968816548181
7.5 0.043458630715622
7.6 0.0493894650833965
7.7 0.0558582399785237
7.8 0.0628145599192114
7.9 0.0701881314185011
8 0.0778890030607453
8.1 0.0858088001305046
8.2 0.0938230072390744
8.3 0.101794287550167
8.4 0.109576755467709
8.5 0.117021046400343
8.6 0.123979958581722
8.7 0.130314384207402
8.8 0.135899206202609
8.9 0.140628817527439
9 0.144421925167791
9.1 0.147225331880127
9.2 0.149016444077617
9.3 0.149804330394624
9.4 0.14962924684813
9.5 0.148560643993916
9.6 0.146693771047118
9.7 0.14414508349635
9.8 0.141046736868851
9.9 0.13754050398058
10 0.133771482168547
10.1 0.129881958897725
10.2 0.126005779415394
10.3 0.122263511708099
10.4 0.118758636712845
10.5 0.115574911715614
10.6 0.112774969027659
10.7 0.110400127307641
10.8 0.10847131567223
10.9 0.10699094630234
11 0.105945523448844
11.1 0.105308747764176
11.2 0.105044865254633
11.3 0.105112018835347
11.4 0.105465385151228
11.5 0.106059916710931
11.6 0.106852555577988
11.7 0.107803835757139
11.8 0.108878842997129
11.9 0.110047549385489
12 0.111284582816964
12.1 0.11256852588974
12.2 0.113880863533688
12.3 0.115204713009705
12.4 0.116523473889476
12.5 0.117819529933515
12.6 0.119073120669048
12.7 0.120261479566597
12.8 0.121358309926014
12.9 0.122333640936517
13 0.123154076896153
13.1 0.123783424171616
13.2 0.124183654840441
13.3 0.124316144479657
13.4 0.124143105289901
13.5 0.123629125326134
13.6 0.122742720301545
13.7 0.121457806110127
13.8 0.119755007394112
13.9 0.117622729383815
14 0.115057935846179
14.1 0.11206659412624
14.2 0.108663767711624
14.3 0.104873356261162
14.4 0.100727501465635
14.5 0.0962656934510462
14.6 0.091533625892061
14.7 0.0865818580086727
14.8 0.081464347857279
14.9 0.0762369237327858
15 0.0709557592392393
15.1 0.0656759130337727
15.2 0.0604499869350909
15.3 0.0553269466599387
15.4 0.0503511386186682
15.5 0.0455615246867059
15.6 0.0409911453620093
15.7 0.0366668108368998
15.8 0.0326090097710009
15.9 0.0288320173426987
16 0.0253441777391238
16.1 0.0221483317457033
16.2 0.019242357517866
16.3 0.0166197918524651
16.4 0.0142705001294211
16.5 0.0121813653046636
16.6 0.0103369696031922
16.7 0.00872024656915045
16.8 0.00731308556651588
16.9 0.00609687540076866
17 0.00505297819701558
17.1 0.0041631288179513
17.2 0.00340975878113854
17.3 0.00277624673647286
17.4 0.00224710003748212
17.5 0.00180807377404012
17.6 0.00144623485535057
17.7 0.00114997939508896
17.8 0.000909011829516103
17.9 0.000714293979926598
18 0.000557971742999746
18.1 0.000433286344703339
18.2 0.000334476207017596
18.3 0.000256674523628496
18.4 0.000195806680670079
18.5 0.000148490738569156
18.6 0.00011194334541859
18.7 8.389270369001e-05
18.8 6.2499572789433e-05
18.9 4.62867636685821e-05
19 3.40771654013743e-05
19.1 2.49400293696352e-05
19.2 1.81450132754489e-05
19.3 1.31233416131841e-05
19.4 9.4353578891498e-06
19.5 6.74371348085574e-06
19.6 4.79144626173621e-06
19.7 3.38423805548825e-06
19.8 2.3761943189358e-06
19.9 1.6585545576127e-06
};
\addlegendentry{GMM fitting};
\end{axis}

\end{tikzpicture}}
		\caption{$X|s_j$ and GMM fitting}
	\end{subfigure}
	\begin{subfigure}{0.33\linewidth}
		\scalebox{0.77}{
\begin{tikzpicture}

\definecolor{darkgray176}{RGB}{176,176,176}
\definecolor{steelblue31119180}{RGB}{185,135,145}

\begin{axis}[
width=3in,
height=2.3in,
tick align=outside,
tick pos=left,
x grid style={darkgray176},
xmin=-2, xmax=22,
xtick style={color=black},
y grid style={darkgray176},
ymin=0, ymax=0.15,
ytick style={color=black},
yticklabel style={
        /pgf/number format/fixed,
        /pgf/number format/precision=3
},
legend style={at={(0.42,0.97)},draw=darkgray!60!black,fill=white,legend cell align=left,nodes={scale=0.7, transform shape}}
]
\draw[draw=none,fill=steelblue31119180] (axis cs:-5,0) rectangle (axis cs:-4,0);
\draw[draw=none,fill=steelblue31119180] (axis cs:-4,0) rectangle (axis cs:-3,0);
\draw[draw=none,fill=steelblue31119180] (axis cs:-3,0) rectangle (axis cs:-2,0.00192492781520693);
\draw[draw=none,fill=steelblue31119180] (axis cs:-2,0) rectangle (axis cs:-1,0.00288739172281039);
\draw[draw=none,fill=steelblue31119180] (axis cs:-1,0) rectangle (axis cs:0,0.00192492781520693);
\draw[draw=none,fill=steelblue31119180] (axis cs:0,0) rectangle (axis cs:1,0.00288739172281039);
\draw[draw=none,fill=steelblue31119180] (axis cs:1,0) rectangle (axis cs:2,0.00192492781520693);
\draw[draw=none,fill=steelblue31119180] (axis cs:2,0) rectangle (axis cs:3,0.0115495668912416);
\draw[draw=none,fill=steelblue31119180] (axis cs:3,0) rectangle (axis cs:4,0.014436958614052);
\draw[draw=none,fill=steelblue31119180] (axis cs:4,0) rectangle (axis cs:5,0.0163618864292589);
\draw[draw=none,fill=steelblue31119180] (axis cs:5,0) rectangle (axis cs:6,0.0211742059672762);
\draw[draw=none,fill=steelblue31119180] (axis cs:6,0) rectangle (axis cs:7,0.0288739172281039);
\draw[draw=none,fill=steelblue31119180] (axis cs:7,0) rectangle (axis cs:8,0.0529355149181906);
\draw[draw=none,fill=steelblue31119180] (axis cs:8,0) rectangle (axis cs:9,0.0981713185755534);
\draw[draw=none,fill=steelblue31119180] (axis cs:9,0) rectangle (axis cs:10,0.124157844080847);
\draw[draw=none,fill=steelblue31119180] (axis cs:10,0) rectangle (axis cs:11,0.115495668912416);
\draw[draw=none,fill=steelblue31119180] (axis cs:11,0) rectangle (axis cs:12,0.0933589990375361);
\draw[draw=none,fill=steelblue31119180] (axis cs:12,0) rectangle (axis cs:13,0.107795957651588);
\draw[draw=none,fill=steelblue31119180] (axis cs:13,0) rectangle (axis cs:14,0.12223291626564);
\draw[draw=none,fill=steelblue31119180] (axis cs:14,0) rectangle (axis cs:15,0.0673724735322425);
\draw[draw=none,fill=steelblue31119180] (axis cs:15,0) rectangle (axis cs:16,0.0490856592877767);
\draw[draw=none,fill=steelblue31119180] (axis cs:16,0) rectangle (axis cs:17,0.0307988450433109);
\draw[draw=none,fill=steelblue31119180] (axis cs:17,0) rectangle (axis cs:18,0.0221366698748797);
\draw[draw=none,fill=steelblue31119180] (axis cs:18,0) rectangle (axis cs:19,0.00673724735322425);
\draw[draw=none,fill=steelblue31119180] (axis cs:19,0) rectangle (axis cs:20,0.00481231953801732);
\draw[draw=none,fill=steelblue31119180] (axis cs:20,0) rectangle (axis cs:21,0.000962463907603465);
\draw[draw=none,fill=steelblue31119180] (axis cs:21,0) rectangle (axis cs:22,0);
\draw[draw=none,fill=steelblue31119180] (axis cs:22,0) rectangle (axis cs:23,0);
\draw[draw=none,fill=steelblue31119180] (axis cs:23,0) rectangle (axis cs:24,0);
\draw[draw=none,fill=steelblue31119180] (axis cs:24,0) rectangle (axis cs:25,0);
\addlegendimage{area legend,fill=steelblue31119180,draw = white}
\addlegendentry[align=left]{noised data \\ $\epsilon=1$, $\delta=0.5$}
\end{axis}

\end{tikzpicture}}
		\caption{$Y|s_j$: $\epsilon = 1$ and $\delta=0.5$}
	\end{subfigure}
	\begin{subfigure}{0.33\linewidth}
		\scalebox{0.77}{
\begin{tikzpicture}

\definecolor{darkgray176}{RGB}{176,176,176}
\definecolor{steelblue31119180}{RGB}{185,135,145}

\begin{axis}[
width=3in,
height=2.3in,
tick align=outside,
tick pos=left,
x grid style={darkgray176},
xmin=-2, xmax=22,
xtick style={color=black},
y grid style={darkgray176},
ymin=0, ymax=0.118550724637681,
ytick style={color=black},
yticklabel style={
        /pgf/number format/fixed,
        /pgf/number format/precision=3
},
legend style={at={(0.42,0.97)},draw=darkgray!60!black,fill=white,legend cell align=left,nodes={scale=0.7, transform shape}}
]
\draw[draw=none,fill=steelblue31119180] (axis cs:-5,0) rectangle (axis cs:-4,0);
\draw[draw=none,fill=steelblue31119180] (axis cs:-4,0) rectangle (axis cs:-3,0.00193236714975845);
\draw[draw=none,fill=steelblue31119180] (axis cs:-3,0) rectangle (axis cs:-2,0.00386473429951691);
\draw[draw=none,fill=steelblue31119180] (axis cs:-2,0) rectangle (axis cs:-1,0.00676328502415459);
\draw[draw=none,fill=steelblue31119180] (axis cs:-1,0) rectangle (axis cs:0,0.000966183574879227);
\draw[draw=none,fill=steelblue31119180] (axis cs:0,0) rectangle (axis cs:1,0.00772946859903382);
\draw[draw=none,fill=steelblue31119180] (axis cs:1,0) rectangle (axis cs:2,0.0106280193236715);
\draw[draw=none,fill=steelblue31119180] (axis cs:2,0) rectangle (axis cs:3,0.00869565217391304);
\draw[draw=none,fill=steelblue31119180] (axis cs:3,0) rectangle (axis cs:4,0.0222222222222222);
\draw[draw=none,fill=steelblue31119180] (axis cs:4,0) rectangle (axis cs:5,0.0193236714975845);
\draw[draw=none,fill=steelblue31119180] (axis cs:5,0) rectangle (axis cs:6,0.029951690821256);
\draw[draw=none,fill=steelblue31119180] (axis cs:6,0) rectangle (axis cs:7,0.0608695652173913);
\draw[draw=none,fill=steelblue31119180] (axis cs:7,0) rectangle (axis cs:8,0.0531400966183575);
\draw[draw=none,fill=steelblue31119180] (axis cs:8,0) rectangle (axis cs:9,0.0714975845410628);
\draw[draw=none,fill=steelblue31119180] (axis cs:9,0) rectangle (axis cs:10,0.0975845410628019);
\draw[draw=none,fill=steelblue31119180] (axis cs:10,0) rectangle (axis cs:11,0.103381642512077);
\draw[draw=none,fill=steelblue31119180] (axis cs:11,0) rectangle (axis cs:12,0.0985507246376812);
\draw[draw=none,fill=steelblue31119180] (axis cs:12,0) rectangle (axis cs:13,0.085024154589372);
\draw[draw=none,fill=steelblue31119180] (axis cs:13,0) rectangle (axis cs:14,0.0985507246376812);
\draw[draw=none,fill=steelblue31119180] (axis cs:14,0) rectangle (axis cs:15,0.0734299516908213);
\draw[draw=none,fill=steelblue31119180] (axis cs:15,0) rectangle (axis cs:16,0.0502415458937198);
\draw[draw=none,fill=steelblue31119180] (axis cs:16,0) rectangle (axis cs:17,0.0270531400966184);
\draw[draw=none,fill=steelblue31119180] (axis cs:17,0) rectangle (axis cs:18,0.021256038647343);
\draw[draw=none,fill=steelblue31119180] (axis cs:18,0) rectangle (axis cs:19,0.0144927536231884);
\draw[draw=none,fill=steelblue31119180] (axis cs:19,0) rectangle (axis cs:20,0.0173913043478261);
\draw[draw=none,fill=steelblue31119180] (axis cs:20,0) rectangle (axis cs:21,0.00579710144927536);
\draw[draw=none,fill=steelblue31119180] (axis cs:21,0) rectangle (axis cs:22,0.00483091787439614);
\draw[draw=none,fill=steelblue31119180] (axis cs:22,0) rectangle (axis cs:23,0.00386473429951691);
\draw[draw=none,fill=steelblue31119180] (axis cs:23,0) rectangle (axis cs:24,0.000966183574879227);
\draw[draw=none,fill=steelblue31119180] (axis cs:24,0) rectangle (axis cs:25,0);
\addlegendimage{area legend,fill=steelblue31119180,draw = white}
\addlegendentry[align=left]{noised data \\ $\epsilon=1$, $\delta=0.3$}
\end{axis}

\end{tikzpicture}}
		\caption{$Y|s_j$: $\epsilon = 1$ and $\delta=0.3$}
	\end{subfigure}
	\caption{The \texttt{Adult} dataset in UCI machine learning repository \cite{UCI2007}: $X$ and $S$ denote the attributes \texttt{education-num} and \texttt{race}, respectively. To attain the statistical indistinguishability between secrets $s_i=$``\texttt{race} is \texttt{Black}" and $s_j=$``\texttt{race} is \texttt{Asian-Pac-Islander}", the privatized data $Y = X + N$ is generated, where in Laplace noise $N\sim\Lap(b)$ is calibrated by Theorem~\ref{theo:LapGaussMix} based on the GMM fitting for attaining $(1,0.5)$-pufferfish privacy and $(1,0.3)$-pufferfish privacy. }
	\label{fig:UCIexp}
\end{figure*}
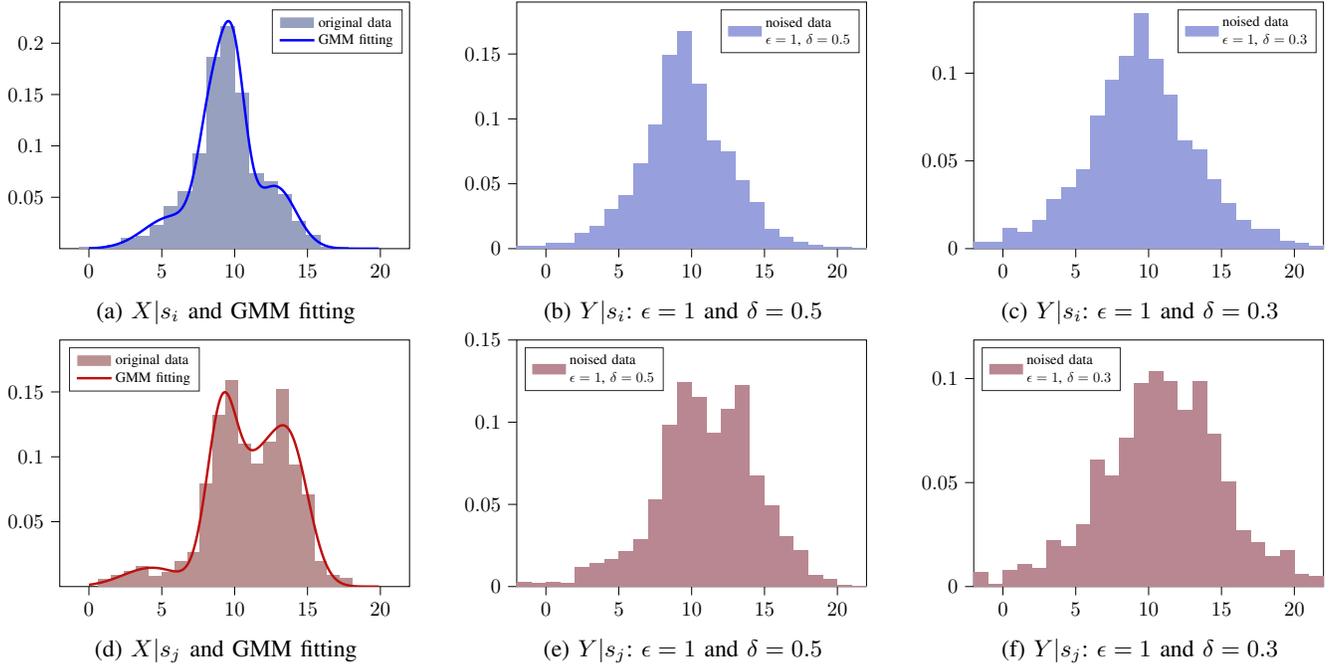

\section{GMM PRIORS}
\label{sec:LapGMM}
For arbitrary distributed $X|s_i$, assume the adversary trains a Gaussian mixture model (GMM) $X|s_i \sim \GaussMix(D_i)$ that is constituted by a finite number $D_i$ of Gaussian components.
That is, under the adversary's prior belief $\rho$, the probability distribution of $X$ given secret $s_i$ is
\begin{equation*}
		P_{X|S}(x|s_i,\rho)  = \sum_{m = 1}^{D_i} \alpha_{im} \GaussProb(x ;  \mu_{im} \sigma_{im}^2),
\end{equation*}	
where $\sum_{m = 1}^{D_i} \alpha_{im} = 1 $ and $\GaussProb(\cdot; \mu_{im}, \sigma_{im}^2)$ denotes the $m$th Gaussian component with mean $\mu_{im}$ and variance $\sigma_{im}^2$.

We apply the optimal transport plan between two GMMs recently derived in \cite{Chen2019GMM,Delon2020GMM}. For $X|s_i \sim \GaussMix(D_i)$ and $X|s_j \sim  \GaussMix(D_j)$, the transport plan $\hat{\pi}$ for $W_2$ distance is\footnote{The transport plan in \eqref{eq:MongeGMM} gives rise to $W_2$ distance if the infimum is taken over all couplings that are GMMs, i.e., it provides an upper bound on the actual $W_2$. See \cite[Sec.~4]{Delon2020GMM}.}
\begin{equation} \label{eq:MongeGMM}
	\hat{\pi} (x,x') = \sum_{m,l} w_{kl}^* \GaussProb(x; \mu_{im}, \sigma_{im}^2) \mathbb{I}\Set{x' = T_{ml}(x)}
\end{equation}
where $T_{ml}$ is the linear mapping \eqref{eq:Tmap} between the $m$th component in $\GaussMix(D_i)$ and $l$th component in $\GaussMix(D_j)$ and $w_{ml}^*, \forall m, l$ is the minimizer of  linear programming $\min \sum_{ml} w_{ml} W_2^2(\GaussProb(\cdot; \mu_{im},\sigma_{im}^2), \GaussProb(\cdot;\mu_{jl},\sigma_{jl}^2))$.\footnote{$W_2^2(\GaussProb(\cdot; \mu_{im},\sigma_{im}^2),\GaussProb(\cdot;\mu_{jl},\sigma_{jl}^2)) = (\mu_{im} - \mu_{jl})^2 + (\sigma_{im} - \sigma_{jl})^2 $ denotes the square of $W_2$ distance between $m$th component in $\GaussMix(D_i)$ and $l$th component in $\GaussMix(D_j)$~\cite[Sec.~2.4.1]{Delon2020GMM}. That is, the optimal transport plan in \eqref{eq:MongeGMM} is fully determined by the parameters of GMMs.}

We show in the following theorem that it suffices to calibrate the noise to the weighted sum of means and variance for GMM priors.

\begin{theorem}\label{theo:LapGaussMix}
	For $X|s_i \sim \GaussMix(D_i)$ and $X|s_j \sim  \GaussMix(D_j)$ for all $(s_i,s_j) \in \SetPair$, adding Laplace noise $N \sim \Lap (b) $ with\footnote{Note that there is an optimal weight $w_{kl}^*$ for each pair of secret $(s_i, s_j) \in \SetPair$.}
	\begin{multline}\label{eq:theo:LapGaussMix}
		b \geq
		\frac{1}{\epsilon}  \max_{\rho,(s_i, s_j)\in \SetPair} \sum_{m,l} w_{ml}^*  \Big( |\mu_{im}-\mu_{jl} | + \\
\tau^*(\delta)|\sigma_{im}-\sigma_{jl}| \Big).
	\end{multline}
	attains $(\epsilon, \delta)$-pufferfish privacy on $\SetPair$ in $Y$.
\end{theorem}

Consider a special case when all GMMs have the same number of components and differ in mean only.  That is,
\begin{equation} \label{eq:GMMmeandiff}
    P_{X|S}(x|s_i) = \sum_{m=1}^{D} \alpha_{m}  \GaussProb(x; \mu_{im}, \sigma_m^2)
\end{equation}
for all secrets $s_i$, where $\mu_{im} \neq  \mu _{jm} $ for each secret pair $(s_i, s_j) \in \SetPair$.
This is a common situation in audio pattern recognition, the so-called GMM-UBM method \cite{GMMUBM2004}: train a universal background model (UBM); then, adapt to the GMM for each individual by changing only the means of Gaussian components in UBM.
It is a maximum a posteriori (MAP) approach, where variance adaptation does not improve the performance of estimation and therefore remains unchanged.

GMM-UBM would very likely be the inference method adopted by an adversary, who is maliciously estimating the arbitrarily distributed statistics for each secret $s_i$.
In this case, for each secret pair $(s_i,s_j)$, all Gaussian components are translations to each other: for each component $m$, $\GaussProb(x; \mu_{im},\sigma_{m}^2) = \GaussProb(x'; \mu_{jm},\sigma_{m}^2), \forall x' = x - \mu_{im} + \mu_{jm}$.
In this case, the following corollary shows that it suffices to only scale $b$ to a convex combination of differences in mean.

\begin{corollary}\label{coro:TransPriorGMM}
	If $X|s_i \sim \GaussMix(D)$ with probability distribution \eqref{eq:GMMmeandiff} for all $s_i$, $Y = X + N$ for $N \sim \Lap (b)$ is $\epsilon$-pufferfish private if
	\begin{equation} \label{eq:TranslationGMM}
		b \geq \frac{1}{\epsilon} \max_{\rho,(s_i, s_j)  \in \SetPair}  \sum_{m} \alpha_m  |\mu_{im} - \mu_{jm}| .
	\end{equation}
\end{corollary}

\begin{figure*}[h]
		\begin{subfigure}{0.33\linewidth}
		\scalebox{0.77}{\input{figures/pdfXgSi_Heart.tex}}
		\caption{$X|s_i$ and GMM fitting}
	\end{subfigure}
	\begin{subfigure}{0.33\linewidth}
		\scalebox{0.77}{
\begin{tikzpicture}

\definecolor{darkgray176}{RGB}{176,176,176}
\definecolor{steelblue31119180}{RGB}{151,160,220}

\begin{axis}[
	width=3in,
	height=2.3in,
tick align=outside,
tick pos=left,
x grid style={darkgray176},
xmin=54, xmax=626,
xtick style={color=black},
y grid style={darkgray176},
ymin=0, ymax=0.00598290598290597,
ytick style={color=black},
legend style={at={(0.98,0.97)},draw=darkgray!60!black,fill=white,legend cell align=left,nodes={scale=0.7, transform shape}},
]
\draw[draw=none,fill=steelblue31119180] (axis cs:80,0) rectangle (axis cs:97.3333333333333,0.00142450142450143);
\draw[draw=none,fill=steelblue31119180] (axis cs:97.3333333333333,0) rectangle (axis cs:114.666666666667,0.00427350427350427);
\draw[draw=none,fill=steelblue31119180] (axis cs:114.666666666667,0) rectangle (axis cs:132,0.00213675213675214);
\draw[draw=none,fill=steelblue31119180] (axis cs:132,0) rectangle (axis cs:149.333333333333,0.00356125356125357);
\draw[draw=none,fill=steelblue31119180] (axis cs:149.333333333333,0) rectangle (axis cs:166.666666666667,0.00427350427350427);
\draw[draw=none,fill=steelblue31119180] (axis cs:166.666666666667,0) rectangle (axis cs:184,0.00356125356125356);
\draw[draw=none,fill=steelblue31119180] (axis cs:184,0) rectangle (axis cs:201.333333333333,0.00356125356125357);
\draw[draw=none,fill=steelblue31119180] (axis cs:201.333333333333,0) rectangle (axis cs:218.666666666667,0.00213675213675214);
\draw[draw=none,fill=steelblue31119180] (axis cs:218.666666666667,0) rectangle (axis cs:236,0.00213675213675214);
\draw[draw=none,fill=steelblue31119180] (axis cs:236,0) rectangle (axis cs:253.333333333333,0.00498575498575499);
\draw[draw=none,fill=steelblue31119180] (axis cs:253.333333333333,0) rectangle (axis cs:270.666666666667,0.00356125356125357);
\draw[draw=none,fill=steelblue31119180] (axis cs:270.666666666667,0) rectangle (axis cs:288,0.00569800569800569);
\draw[draw=none,fill=steelblue31119180] (axis cs:288,0) rectangle (axis cs:305.333333333333,0.00498575498575499);
\draw[draw=none,fill=steelblue31119180] (axis cs:305.333333333333,0) rectangle (axis cs:322.666666666667,0.000712250712250713);
\draw[draw=none,fill=steelblue31119180] (axis cs:322.666666666667,0) rectangle (axis cs:340,0.00356125356125355);
\draw[draw=none,fill=steelblue31119180] (axis cs:340,0) rectangle (axis cs:357.333333333333,0.00427350427350428);
\draw[draw=none,fill=steelblue31119180] (axis cs:357.333333333333,0) rectangle (axis cs:374.666666666667,0);
\draw[draw=none,fill=steelblue31119180] (axis cs:374.666666666667,0) rectangle (axis cs:392,0);
\draw[draw=none,fill=steelblue31119180] (axis cs:392,0) rectangle (axis cs:409.333333333333,0.000712250712250713);
\draw[draw=none,fill=steelblue31119180] (axis cs:409.333333333333,0) rectangle (axis cs:426.666666666667,0.000712250712250713);
\draw[draw=none,fill=steelblue31119180] (axis cs:426.666666666667,0) rectangle (axis cs:444,0);
\draw[draw=none,fill=steelblue31119180] (axis cs:444,0) rectangle (axis cs:461.333333333333,0.000712250712250713);
\draw[draw=none,fill=steelblue31119180] (axis cs:461.333333333333,0) rectangle (axis cs:478.666666666667,0);
\draw[draw=none,fill=steelblue31119180] (axis cs:478.666666666667,0) rectangle (axis cs:496,0);
\draw[draw=none,fill=steelblue31119180] (axis cs:496,0) rectangle (axis cs:513.333333333333,0);
\draw[draw=none,fill=steelblue31119180] (axis cs:513.333333333333,0) rectangle (axis cs:530.666666666667,0);
\draw[draw=none,fill=steelblue31119180] (axis cs:530.666666666667,0) rectangle (axis cs:548,0.000712250712250711);
\draw[draw=none,fill=steelblue31119180] (axis cs:548,0) rectangle (axis cs:565.333333333333,0);
\draw[draw=none,fill=steelblue31119180] (axis cs:565.333333333333,0) rectangle (axis cs:582.666666666667,0);
\draw[draw=none,fill=steelblue31119180] (axis cs:582.666666666667,0) rectangle (axis cs:600,0);
\addlegendimage{area legend,fill=steelblue31119180,draw = white}
\addlegendentry[align=left]{noised data \\ $\epsilon=1$, $\delta=0.5$}
\end{axis}

\end{tikzpicture}}
		\caption{$Y|s_i$: $\epsilon = 1$ and $\delta=0.5$}
	\end{subfigure}
	\begin{subfigure}{0.33\linewidth}
		\scalebox{0.77}{
\begin{tikzpicture}

\definecolor{darkgray176}{RGB}{176,176,176}
\definecolor{steelblue31119180}{RGB}{151,160,220}

\begin{axis}[
	width=3in,
	height=2.3in,
tick align=outside,
tick pos=left,
x grid style={darkgray176},
xmin=54, xmax=626,
xtick style={color=black},
y grid style={darkgray176},
ymin=0, ymax=0.00565384615384616,
ytick style={color=black},
legend style={at={(0.98,0.97)},draw=darkgray!60!black,fill=white,legend cell align=left,nodes={scale=0.7, transform shape}},
]
\draw[draw=none,fill=steelblue31119180] (axis cs:80,0) rectangle (axis cs:97.3333333333333,0.00230769230769231);
\draw[draw=none,fill=steelblue31119180] (axis cs:97.3333333333333,0) rectangle (axis cs:114.666666666667,0.00230769230769231);
\draw[draw=none,fill=steelblue31119180] (axis cs:114.666666666667,0) rectangle (axis cs:132,0.00153846153846154);
\draw[draw=none,fill=steelblue31119180] (axis cs:132,0) rectangle (axis cs:149.333333333333,0.00230769230769231);
\draw[draw=none,fill=steelblue31119180] (axis cs:149.333333333333,0) rectangle (axis cs:166.666666666667,0.00230769230769231);
\draw[draw=none,fill=steelblue31119180] (axis cs:166.666666666667,0) rectangle (axis cs:184,0.00384615384615384);
\draw[draw=none,fill=steelblue31119180] (axis cs:184,0) rectangle (axis cs:201.333333333333,0.00307692307692308);
\draw[draw=none,fill=steelblue31119180] (axis cs:201.333333333333,0) rectangle (axis cs:218.666666666667,0.00461538461538461);
\draw[draw=none,fill=steelblue31119180] (axis cs:218.666666666667,0) rectangle (axis cs:236,0.00538461538461538);
\draw[draw=none,fill=steelblue31119180] (axis cs:236,0) rectangle (axis cs:253.333333333333,0.00307692307692308);
\draw[draw=none,fill=steelblue31119180] (axis cs:253.333333333333,0) rectangle (axis cs:270.666666666667,0.00538461538461539);
\draw[draw=none,fill=steelblue31119180] (axis cs:270.666666666667,0) rectangle (axis cs:288,0.00307692307692307);
\draw[draw=none,fill=steelblue31119180] (axis cs:288,0) rectangle (axis cs:305.333333333333,0.00230769230769231);
\draw[draw=none,fill=steelblue31119180] (axis cs:305.333333333333,0) rectangle (axis cs:322.666666666667,0.00461538461538462);
\draw[draw=none,fill=steelblue31119180] (axis cs:322.666666666667,0) rectangle (axis cs:340,0.00307692307692307);
\draw[draw=none,fill=steelblue31119180] (axis cs:340,0) rectangle (axis cs:357.333333333333,0);
\draw[draw=none,fill=steelblue31119180] (axis cs:357.333333333333,0) rectangle (axis cs:374.666666666667,0.00307692307692308);
\draw[draw=none,fill=steelblue31119180] (axis cs:374.666666666667,0) rectangle (axis cs:392,0);
\draw[draw=none,fill=steelblue31119180] (axis cs:392,0) rectangle (axis cs:409.333333333333,0.00230769230769231);
\draw[draw=none,fill=steelblue31119180] (axis cs:409.333333333333,0) rectangle (axis cs:426.666666666667,0.00076923076923077);
\draw[draw=none,fill=steelblue31119180] (axis cs:426.666666666667,0) rectangle (axis cs:444,0.000769230769230767);
\draw[draw=none,fill=steelblue31119180] (axis cs:444,0) rectangle (axis cs:461.333333333333,0);
\draw[draw=none,fill=steelblue31119180] (axis cs:461.333333333333,0) rectangle (axis cs:478.666666666667,0);
\draw[draw=none,fill=steelblue31119180] (axis cs:478.666666666667,0) rectangle (axis cs:496,0);
\draw[draw=none,fill=steelblue31119180] (axis cs:496,0) rectangle (axis cs:513.333333333333,0);
\draw[draw=none,fill=steelblue31119180] (axis cs:513.333333333333,0) rectangle (axis cs:530.666666666667,0.00153846153846153);
\draw[draw=none,fill=steelblue31119180] (axis cs:530.666666666667,0) rectangle (axis cs:548,0);
\draw[draw=none,fill=steelblue31119180] (axis cs:548,0) rectangle (axis cs:565.333333333333,0);
\draw[draw=none,fill=steelblue31119180] (axis cs:565.333333333333,0) rectangle (axis cs:582.666666666667,0);
\draw[draw=none,fill=steelblue31119180] (axis cs:582.666666666667,0) rectangle (axis cs:600,0);
\addlegendimage{area legend,fill=steelblue31119180,draw = white}
\addlegendentry[align=left]{noised data \\ $\epsilon=1$, $\delta=0.3$}
\end{axis}
\end{tikzpicture}}
		\caption{$Y|s_i$: $\epsilon = 1$ and $\delta=0.3$}
	\end{subfigure} \\
	\begin{subfigure}{0.33\linewidth}
		\scalebox{0.77}{\input{figures/pdfXgSj_Heart.tex}}
		\caption{$X|s_j$ and GMM fitting}
	\end{subfigure}
	\begin{subfigure}{0.33\linewidth}
		\scalebox{0.77}{
\begin{tikzpicture}

\definecolor{darkgray176}{RGB}{176,176,176}
\definecolor{steelblue31119180}{RGB}{185,135,145}

\begin{axis}[
	width=3in,
	height=2.3in,
tick align=outside,
tick pos=left,
x grid style={darkgray176},
xmin=54, xmax=626,
xtick style={color=black},
y grid style={darkgray176},
ymin=0, ymax=0.00634615384615385,
ytick style={color=black},
legend style={at={(0.98,0.97)},draw=darkgray!60!black,fill=white,legend cell align=left,nodes={scale=0.7, transform shape}},
]
\draw[draw=none,fill=steelblue31119180] (axis cs:80,0) rectangle (axis cs:97.3333333333333,0.000824175824175824);
\draw[draw=none,fill=steelblue31119180] (axis cs:97.3333333333333,0) rectangle (axis cs:114.666666666667,0.0010989010989011);
\draw[draw=none,fill=steelblue31119180] (axis cs:114.666666666667,0) rectangle (axis cs:132,0.00247252747252747);
\draw[draw=none,fill=steelblue31119180] (axis cs:132,0) rectangle (axis cs:149.333333333333,0.00192307692307693);
\draw[draw=none,fill=steelblue31119180] (axis cs:149.333333333333,0) rectangle (axis cs:166.666666666667,0.00247252747252747);
\draw[draw=none,fill=steelblue31119180] (axis cs:166.666666666667,0) rectangle (axis cs:184,0.0021978021978022);
\draw[draw=none,fill=steelblue31119180] (axis cs:184,0) rectangle (axis cs:201.333333333333,0.00604395604395605);
\draw[draw=none,fill=steelblue31119180] (axis cs:201.333333333333,0) rectangle (axis cs:218.666666666667,0.00604395604395604);
\draw[draw=none,fill=steelblue31119180] (axis cs:218.666666666667,0) rectangle (axis cs:236,0.00439560439560439);
\draw[draw=none,fill=steelblue31119180] (axis cs:236,0) rectangle (axis cs:253.333333333333,0.0043956043956044);
\draw[draw=none,fill=steelblue31119180] (axis cs:253.333333333333,0) rectangle (axis cs:270.666666666667,0.00521978021978023);
\draw[draw=none,fill=steelblue31119180] (axis cs:270.666666666667,0) rectangle (axis cs:288,0.00494505494505493);
\draw[draw=none,fill=steelblue31119180] (axis cs:288,0) rectangle (axis cs:305.333333333333,0.00412087912087913);
\draw[draw=none,fill=steelblue31119180] (axis cs:305.333333333333,0) rectangle (axis cs:322.666666666667,0.00247252747252748);
\draw[draw=none,fill=steelblue31119180] (axis cs:322.666666666667,0) rectangle (axis cs:340,0.00192307692307692);
\draw[draw=none,fill=steelblue31119180] (axis cs:340,0) rectangle (axis cs:357.333333333333,0.00192307692307693);
\draw[draw=none,fill=steelblue31119180] (axis cs:357.333333333333,0) rectangle (axis cs:374.666666666667,0.000824175824175825);
\draw[draw=none,fill=steelblue31119180] (axis cs:374.666666666667,0) rectangle (axis cs:392,0.0010989010989011);
\draw[draw=none,fill=steelblue31119180] (axis cs:392,0) rectangle (axis cs:409.333333333333,0);
\draw[draw=none,fill=steelblue31119180] (axis cs:409.333333333333,0) rectangle (axis cs:426.666666666667,0.000824175824175825);
\draw[draw=none,fill=steelblue31119180] (axis cs:426.666666666667,0) rectangle (axis cs:444,0);
\draw[draw=none,fill=steelblue31119180] (axis cs:444,0) rectangle (axis cs:461.333333333333,0.0010989010989011);
\draw[draw=none,fill=steelblue31119180] (axis cs:461.333333333333,0) rectangle (axis cs:478.666666666667,0.000274725274725275);
\draw[draw=none,fill=steelblue31119180] (axis cs:478.666666666667,0) rectangle (axis cs:496,0.000274725274725274);
\draw[draw=none,fill=steelblue31119180] (axis cs:496,0) rectangle (axis cs:513.333333333333,0.000274725274725276);
\draw[draw=none,fill=steelblue31119180] (axis cs:513.333333333333,0) rectangle (axis cs:530.666666666667,0);
\draw[draw=none,fill=steelblue31119180] (axis cs:530.666666666667,0) rectangle (axis cs:548,0.000274725274725274);
\draw[draw=none,fill=steelblue31119180] (axis cs:548,0) rectangle (axis cs:565.333333333333,0);
\draw[draw=none,fill=steelblue31119180] (axis cs:565.333333333333,0) rectangle (axis cs:582.666666666667,0);
\draw[draw=none,fill=steelblue31119180] (axis cs:582.666666666667,0) rectangle (axis cs:600,0.000274725274725274);
\addlegendimage{area legend,fill=steelblue31119180,draw = white}
\addlegendentry[align=left]{noised data \\ $\epsilon=1$, $\delta=0.5$}
\end{axis}

\end{tikzpicture}}
		\caption{$Y|s_j$: $\epsilon = 1$ and $\delta=0.5$}
	\end{subfigure}
	\begin{subfigure}{0.33\linewidth}
		\scalebox{0.77}{
\begin{tikzpicture}

\definecolor{darkgray176}{RGB}{176,176,176}
\definecolor{steelblue31119180}{RGB}{185,135,145}

\begin{axis}[
	width=3in,
	height=2.3in,
tick align=outside,
tick pos=left,
x grid style={darkgray176},
xmin=54, xmax=626,
xtick style={color=black},
y grid style={darkgray176},
ymin=0, ymax=0.00531894934333958,
ytick style={color=black},
legend style={at={(0.98,0.97)},draw=darkgray!60!black,fill=white,legend cell align=left,nodes={scale=0.7, transform shape}},
]
\draw[draw=none,fill=steelblue31119180] (axis cs:80,0) rectangle (axis cs:97.3333333333333,0.00112570356472796);
\draw[draw=none,fill=steelblue31119180] (axis cs:97.3333333333333,0) rectangle (axis cs:114.666666666667,0.00140712945590994);
\draw[draw=none,fill=steelblue31119180] (axis cs:114.666666666667,0) rectangle (axis cs:132,0.00225140712945591);
\draw[draw=none,fill=steelblue31119180] (axis cs:132,0) rectangle (axis cs:149.333333333333,0.00309568480300188);
\draw[draw=none,fill=steelblue31119180] (axis cs:149.333333333333,0) rectangle (axis cs:166.666666666667,0.00168855534709193);
\draw[draw=none,fill=steelblue31119180] (axis cs:166.666666666667,0) rectangle (axis cs:184,0.00393996247654784);
\draw[draw=none,fill=steelblue31119180] (axis cs:184,0) rectangle (axis cs:201.333333333333,0.00365853658536586);
\draw[draw=none,fill=steelblue31119180] (axis cs:201.333333333333,0) rectangle (axis cs:218.666666666667,0.00281425891181989);
\draw[draw=none,fill=steelblue31119180] (axis cs:218.666666666667,0) rectangle (axis cs:236,0.00506566604127579);
\draw[draw=none,fill=steelblue31119180] (axis cs:236,0) rectangle (axis cs:253.333333333333,0.00478424015009381);
\draw[draw=none,fill=steelblue31119180] (axis cs:253.333333333333,0) rectangle (axis cs:270.666666666667,0.00337711069418387);
\draw[draw=none,fill=steelblue31119180] (axis cs:270.666666666667,0) rectangle (axis cs:288,0.00365853658536585);
\draw[draw=none,fill=steelblue31119180] (axis cs:288,0) rectangle (axis cs:305.333333333333,0.00450281425891183);
\draw[draw=none,fill=steelblue31119180] (axis cs:305.333333333333,0) rectangle (axis cs:322.666666666667,0.00225140712945591);
\draw[draw=none,fill=steelblue31119180] (axis cs:322.666666666667,0) rectangle (axis cs:340,0.00309568480300187);
\draw[draw=none,fill=steelblue31119180] (axis cs:340,0) rectangle (axis cs:357.333333333333,0.00309568480300188);
\draw[draw=none,fill=steelblue31119180] (axis cs:357.333333333333,0) rectangle (axis cs:374.666666666667,0.00140712945590995);
\draw[draw=none,fill=steelblue31119180] (axis cs:374.666666666667,0) rectangle (axis cs:392,0.0022514071294559);
\draw[draw=none,fill=steelblue31119180] (axis cs:392,0) rectangle (axis cs:409.333333333333,0.000562851782363978);
\draw[draw=none,fill=steelblue31119180] (axis cs:409.333333333333,0) rectangle (axis cs:426.666666666667,0.00112570356472796);
\draw[draw=none,fill=steelblue31119180] (axis cs:426.666666666667,0) rectangle (axis cs:444,0);
\draw[draw=none,fill=steelblue31119180] (axis cs:444,0) rectangle (axis cs:461.333333333333,0.000281425891181989);
\draw[draw=none,fill=steelblue31119180] (axis cs:461.333333333333,0) rectangle (axis cs:478.666666666667,0.000562851782363978);
\draw[draw=none,fill=steelblue31119180] (axis cs:478.666666666667,0) rectangle (axis cs:496,0.000844277673545964);
\draw[draw=none,fill=steelblue31119180] (axis cs:496,0) rectangle (axis cs:513.333333333333,0);
\draw[draw=none,fill=steelblue31119180] (axis cs:513.333333333333,0) rectangle (axis cs:530.666666666667,0);
\draw[draw=none,fill=steelblue31119180] (axis cs:530.666666666667,0) rectangle (axis cs:548,0);
\draw[draw=none,fill=steelblue31119180] (axis cs:548,0) rectangle (axis cs:565.333333333333,0.00028142589118199);
\draw[draw=none,fill=steelblue31119180] (axis cs:565.333333333333,0) rectangle (axis cs:582.666666666667,0.000281425891181988);
\draw[draw=none,fill=steelblue31119180] (axis cs:582.666666666667,0) rectangle (axis cs:600,0.000281425891181988);
\addlegendimage{area legend,fill=steelblue31119180,draw = white}
\addlegendentry[align=left]{noised data \\ $\epsilon=1$, $\delta=0.3$}
\end{axis}

\end{tikzpicture}}
		\caption{$Y|s_j$: $\epsilon = 1$ and $\delta=0.3$}
	\end{subfigure}
	\caption{The \texttt{Hungarian heart disease}  dataset in UCI machine learning repository \cite{UCI2007}: $X$ and $S$ denote the attributes \texttt{chol}, the cholesterol  level, and \texttt{sex}, respectively. To attain the statistical indistinguishability between secrets $s_i=$``\texttt{sex} is \texttt{female}" and  $s_j=$``\texttt{sex} is \texttt{male}", the privatized data $Y = X + N$ is generated, where in Laplace noise $N\sim\Lap(b)$ is calibrated by Theorem~\ref{theo:LapGaussMix} based on the GMM fitting for attaining $(1,0.5)$-pufferfish privacy and $(1,0.3)$-pufferfish privacy. }
	\label{fig:UCIexp_Heart}
\end{figure*}

\section{Experiment}
\label{sec:UCIexp}

In the UCI machine learning repository~\cite{UCI2007}, the \texttt{adult} dataset was extracted from the census bureau database containing 32652 instances/individuals/records and 15 attributes.
In this experiment, $S$ refers to the sensitive attribute \texttt{race}, a categorical variable, and $X$ refers to attribute \texttt{education-num}, an integer number indicating an individual's education level.
We simulate a scenario that \texttt{education-num} column is to be published. It is assumed that the adversary have access to all published records of \texttt{education-num} and can statistically infer the information on \texttt{race}.
Therefore, the data curator requires a certain level of statistical indistinguishability (by specifying the values of $\epsilon$ and $\delta$) between the secrets: $s_i=$``\texttt{race} is \texttt{Black}" and $s_j=$``\texttt{race} is \texttt{Asian-Pac-Islander}".

In this case, $X$ refers to the values of \texttt{education-num} in all 32652 records; $X|s_i$ refers to the values of \texttt{education-num} in all records having attribute \texttt{race} being ``\texttt{race} is \texttt{Black}"; $X|s_j$ refers to the values of \texttt{education-num} in all records having attribute \texttt{race} being ``\texttt{race} is \texttt{Asian-Pac-Islander}".\footnote{Here, both $X|s_i$ and $X|s_j$ refer to a set of data records, instead of a query answer, with the probability for each instance $x$ to appear in the dataset governed by $P_{X|S}(x|s_i)$ and $P_{X|S}(x|s_j)$, respectively. Therefore, the DP framework proposed \cite{Dwork2006,CalibNoiseDP} does not fit in this case.}
Noise $N$ is added to $X$ to generate privatized data $Y$ referring to 32652 randomized values of \texttt{education-num}.\footnote{Noise $N$ is assumed to be a continuous random variable and therefore the value of $Y$ is continuously changing.} 
By Definition~\ref{def},
the problem is to ensure $P_{Y|S}(B |s_i, \rho ) \leq e^{\epsilon}P_{Y|S}(B |s_j, \rho) + \delta$ for any real number subset $B$. 
That is, the adversary should have difficulty telling whether ``\texttt{race} is \texttt{Black}" or ``\texttt{race} is \texttt{Asian-Pac-Islander}" by observing the randomized \texttt{education-num}s in $Y$, regardless of the observing order.\footnote{Even if in a typical scenario where the data curator releases the randomized \texttt{education-num} $Y$ for all black people first and then for all Asian-Pac-Islander people,  the change in statistics in  $Y$ is under control.}

We first fit two three-component GMMs to the empirical distributions of $X|s_i$ and $X|s_j$, respectively (see Fig.~\ref{fig:UCIexp}(a) and (d)). 
Here, the GMMs denote the adversary's belief or side information $\rho$ on $X$ given two secrets $s_i$ and $s_j$.
For $\epsilon = 1$ and $\delta = 0.5$, calculating the Laplace parameter $b$ by applying Theorem~\ref{theo:LapGaussMix}, we generate privatized data $Y = X + N$ where $N \sim \Lap(b)$ is calibrated by Theorem~\ref{theo:LapGaussMix}. The plots in Fig.~\ref{fig:UCIexp}(b) and (e) show the statistics of $Y$ given $s_i$ and $s_j$, where the differences in empirical probability is reduced.
Repeating the same procedure for a more strict privacy constraint: $\epsilon = 1$ and $\delta = 0.3$, the statistical indistinguishability in the randomized data is further improved. See the two plots in Fig.~\ref{fig:UCIexp}(c) and (f).
The experiment results show the noise calibration methods proposed in this paper work for any given privacy constraints $\epsilon$ and $\delta$. The exact values of $\epsilon$ and $\delta$ can be determined by data privacy requirements in actual applications.

In addition, we repeat the same experiment using the another dataset in the UCI machine learning repository: the \texttt{Hungarian heart disease} dataset was created by the Hungarian Institute of Cardiology, Budapest, which records $293$ patients' data of $76$ attributes for the purpose of identifying the presence of heart disease.
We  extract two attributes: \texttt{sex} as the sensitive data $S$ and \texttt{chol}, denoting the serum cholesterol in mg/dl, as the public data $X$.
We consider $X$ given two secrets:  $s_i=$``\texttt{sex} is \texttt{female}" and $s_j=$``\texttt{sex} is \texttt{male}".
The results are in Fig.~\ref{fig:UCIexp_Heart}. The same as Fig.~\ref{fig:UCIexp}, they show that one can apply the sufficient condition~\eqref{theo:LapGaussMix} to attain $(\epsilon,\delta)$-pufferfish privacy.

\section{DISCUSSION}

It is understandable that the privacy protection should not severely undermine the useful information in the released data $Y$.
For example, the noised counting query in Section~\ref{sec:CountQ} should prevent the malicious inference on individual's data or existence, but still report the summation with the highest accuracy.
It is noted that the conditions  derived in Theorems~\ref{theo:Lap} and \ref{theo:LapGaussMix} are sufficient only. Yet, there is a problem of how to improve these results to further reduce the data distortion but still guarantee a specific degree of  statistical indistinguishability, i.e., to minimize the noise for $(\epsilon,\delta)$-pufferfish private $Y$.
This section discusses possible solutions: tighten the upper bound on $P_{Y|S}(B|s_i, \rho )  - e^{\epsilon}P_{Y|S}(B|s_j, \rho)$ that points out several directions for future works.

\subsection{Tighter Bound on~\eqref{eq:mainInEqAux}}
The idea of proving Theorems~\ref{theo:Lap} and \ref{theo:LapGaussMix} is to derive an upper bound on~\eqref{eq:mainInEqAux}: $P_{Y|S}(B|s_i, \rho )  - e^{\epsilon}P_{Y|S}(B|s_j, \rho) \leq U$ for all $B \subseteq \Real$ and then request $U \leq \delta$ to work out the value of scale parameter $b$ in Laplace noise for given $\epsilon$ and $\delta$.
See Appendices~\ref{app:theo:Lap} and \ref{app:theo:LapGaussMix}.
It is clear the tighter $U$ is, the smaller value of $b$ can be derived and the smaller the amount of noise added to the $(\epsilon,\delta)$-pufferfish privacy attaining released data.
It is worth studying whether a tighter $U$ can be derived.

\subsubsection{Reducing $\tau^*(\delta)$}
In Theorem~\ref{theo:Lap}, the maximand in the sufficient condition~\eqref{eq:theo:Lap} consists of the differences in mean $|\mu_i-\mu_j|$ and standard deviation $|\sigma_i - \sigma_j|$, where the latter is scaled by $\tau^*(\delta)$.
Clearly, minimizing $\tau^*(\delta)$ that attains $(\epsilon,\delta)$-pufferfish privacy will result in a smaller $b$ and therefore a reduction in noise power.
In this paper, the value of $\tau^*(\delta)$ is derived by a sequence of upper bounds on $\int \big( 1 - e^{\epsilon - \frac{|x-T(x)|}{b}} \big) \dif P_{X|S}(x|s_i,\rho)$. See \eqref{eq:W1Exponential:A} to \eqref{eq:W1Exponential:Zsym} in Appendix~\ref{app:theo:Lap}.
It is possible to tighten these upper bounds or the Gaussian tail bound in \eqref{eq:GaussTail}, e.g., by referring to the analytical tightening method such as \cite{GaussDenoise2018}, to further reduce $\tau^*(\delta)$.

\subsubsection{Transport plan other than $\hat{\pi}$}
Section~\ref{sec:Monge} points out the approach of tightening the bound on~\eqref{eq:mainInEqAux} by a minimization over all couplings $\inf_{\pi} \{ P_{Y|S}(B|s_i,\rho) - e^\epsilon P_{Y|S}(B|s_j,\rho)\} $.
This paper exploits the existing Monge's optimal coupling $\hat{\pi}$ for Gaussian priors, which only searches an upper bound on this infimum (see Appendix~\ref{append:W2}).
It is of interest whether there exist other couplings that can improve Theorems~\ref{theo:Lap} and \ref{theo:LapGaussMix}.

Finally,  the results in this paper can be refined or simplified for some specific settings or applications, e.g., the summation query where individual's random variable $Z_k$ is binary, where one might be able to find better solution than Monge's optimal transport plan or a smaller $\delta^*(\delta)$.

\subsection{Exponential Mechanism}

The Laplace distribution belongs to the exponential family, for which we use the inequality of the exponential function $e^{\epsilon - \frac{|y-T(x)|-|y-x|}{b}} \geq e^{\epsilon - \frac{|x-T(x)|}{b}}$ and upper bound $\int \big( 1 - e^{\epsilon - \frac{|x-T(x)|}{b}} \big) \dif P_{X|S}(x|s_i,\rho) \leq \delta$ thereafter in the proofs of Theorems~1 and 2, in Appendices~\ref{app:theo:Lap} and \ref{app:theo:LapGaussMix}, respectively.
This implies the exponential mechanism with noise probability $P_{N} (z) \propto e^{ -\eta(\theta) d(z)}$ for some metric $d(\cdot)$ can be applied to  approximate the $\epsilon$-pufferfish privacy for Gaussian priors, too.
To do so, one can refer to \cite{CalibNoiseDP}  that extends the noise calibration result on Laplace mechanism to exponential mechanism for attaining $\epsilon$-DP.
Among all the probability distributions in the exponential family, we specifically discuss the Gaussian mechanism below.

\paragraph*{Gaussian Noise}
Gaussian mechanism is another commonly used privatization scheme, which might be favored to enhance data utility.
One reason is that the noise variance proportional to the $\ell_2$-norm is no larger than $\ell_1$-norm. This can be seen from Fig.~\ref{fig:Sample}, showing that with the same noise power, Gaussian mechanism provides higher statistical indistinguishability than Laplace mechanism.
In addition, the tail probability of Gaussian distribution decays faster than Laplace distribution, as pointed out by \cite{RDP2017}, and therefore results in an accurate query answer.

While this paper focuses on Laplace mechanism, we suggest the design of Gaussian mechanism as follows.
For Gaussian noise $N \sim \Gauss(0,\theta^2)$ and Gaussian prior $\rho$, the privatized data $Y = X + N $ conditioned on each secret $s_i$ is necessarily Gaussian distributed with mean $\mu_i$ and variance $\theta^2 + \sigma_i^2$.
Then, the problem of attaining $(\epsilon,\delta)$-pufferfish privacy is to derive a sufficient condition in the form of $\theta^2 \geq \xi(\epsilon, \delta, \mu_i, \mu_j, \sigma_i^2, \sigma_j^2)$ such that
$P_{Y|s}(B|s_i,\rho) - e^{\epsilon} P_{Y|s}(B|s_j,\rho) \leq \delta, \forall B \subseteq \Real$
where $Y|s_i \sim \Gauss(\mu_i, \theta^2 + \sigma_i^2)$ and $Y|s_j \sim \Gauss(\mu_j, \theta^2 + \sigma_j^2)$.
This reduces to determining the sufficient condition on $\sigma_1^2 = \theta^2 + \sigma_i^2$ or $\sigma_2^2 = \theta^2 + \sigma_j^2$ such that \cite{TightDelta2018}
\begin{equation} \label{eq:Tight}
    \int [\GaussProb(y; \mu_1, \sigma_1^2) - e^\epsilon \GaussProb(y; \mu_2, \sigma_2^2)]_+ \dif y \leq \delta,
\end{equation}
where $[z]_+ = \max\Set{z,0}$.
The main difficulty is to obtain an estimate of the left hand side of \eqref{eq:Tight}, for which one can still refer to the proofs of Theorem~\ref{theo:Lap} or the FFT method proposed in \cite{TightDeltaFFT2020}.
Once the Gaussian mechanism is designed, one can compare it experimentally to the Laplace mechanism in~Figs.~\ref{fig:UCIexp} and \ref{fig:UCIexp_Heart} for the same values of $\epsilon$ and $\delta$.

\section{Conclusion}
Considering adding Laplace noise $N \sim \Lap(b)$ to the normally distributed data, we derived a lower bound on the scale parameter $b$ for attaining $(\epsilon, \delta)$-pufferfish privacy in the noised data.
It is shown that the $\epsilon$-indistinguishability is guaranteed if the noise is sufficient enough to compromise the difference in mean and variance conditioned on the discriminative secret pair $(s_i,s_j)$.
An application of this result to the summation query revealed that as the number of participants increases, it is more difficulty for an adversary to differentiate individual participants and therefore requires less noise for attaining pufferfish privacy.
When the noise is pairwisely calibrated to the convex combination of difference in mean and variance of Gaussian components, $(\epsilon, \delta)$-pufferfish privacy is attained for any arbitrarily distributed data that is modeled by GMM.
Finally, we pointed out two ways for improving the results in this paper for a higher data utility.

\appendices
\section{Proof of Theorem~\ref{theo:Lap}}
\label{app:theo:Lap}
For each $(s_i,s_j) \in \SetPair$, we have \eqref{eq:mainInEq} equal to
\begin{align}
      & \int \int_B \frac{1}{2b}  \big( e^{-\frac{|y-x|}{b}} - e^{\epsilon - \frac{|y-T(x)|}{b}} \big) \dif y \dif P_{X|S}(x|s_i,\rho)  \nonumber \\
     & = \int \int_B \frac{1}{2b} e^{-\frac{|y-x|}{b}} \big( 1 - e^{\epsilon - \frac{|y-T(x)|-|y-x|}{b}} \big) \dif y \dif P_{X|S}(x|s_i,\rho)  \nonumber \\
     & \leq \int_A \int_B \frac{1}{2b} e^{-\frac{|y-x|}{b}} \dif y \big( 1 - e^{\epsilon - \frac{|x-T(x)|}{b}} \big) \dif P_{X|S}(x|s_i,\rho)  \label{eq:W1Exponential:TriangIneq}\\
     & \leq \int_A \big( 1 - e^{\epsilon - \frac{|x-T(x)|}{b}} \big) \dif P_{X|S}(x|s_i,\rho) \nonumber \\
     & \leq   P_{X|S}(A|s_i,\rho), \qquad \forall B \subseteq \Real, \label{eq:W1Exponential:A}
\end{align}
where inequality~\eqref{eq:W1Exponential:TriangIneq} is because of the triangular inequality $|y-T(x)| - |y-x| \leq |x-T(x)|$ and $A =  \Set{x \colon |x-T(x)| > \epsilon b }$ in \eqref{eq:W1Exponential:A}.
Assume $b \geq \frac{|\mu_i - \mu_j|+c}{\epsilon}$, where $c \in \RealP$. For $X \sim \Gauss(\mu_i,\sigma_i)$, let $Z = \frac{X-\mu_i}{\sigma_i} \sim \Gauss(0,1)$. We have
\begin{small}
\begin{align}
	P_{X|S}(A|s_i,\rho) &= \Pr( |X- T(X)|  > \epsilon b ) \nonumber \\
	&= 	\Pr\big( \big| X  - \mu_j - \frac{X-\mu_i}{\sigma_i} \sigma_j \big|  > \epsilon b \big)  \nonumber \\
	&= 	\Pr (  | Z (\sigma_i - \sigma_j) + (\mu_i - \mu_j)  |  > \epsilon b  )  \nonumber \\
	&= 	\Pr \Big(  \Big| Z  + \frac{\mu_i - \mu_j}{\sigma_i - \sigma_j} \Big|  > \frac{\epsilon b}{|\sigma_i - \sigma_j|}  \Big) \nonumber\\
    &= \begin{cases}
            \Pr \Big(  \Big| Z  + \frac{|\mu_i - \mu_j|}{|\sigma_i - \sigma_j|} \Big|  > \frac{\epsilon b}{|\sigma_i - \sigma_j|}  \Big) & \frac{\mu_i-\mu_j}{\sigma_i - \sigma_j} \geq 0 \\
            \Pr \Big(  \Big| Z  - \frac{|\mu_i - \mu_j|}{|\sigma_i - \sigma_j|} \Big|  > \frac{\epsilon b}{|\sigma_i - \sigma_j|}  \Big) & \frac{\mu_i-\mu_j}{\sigma_i - \sigma_j} < 0
         \end{cases} \nonumber \\
    &\leq \begin{cases}
            2\Pr \big( Z > \frac{\epsilon b - |\mu_i - \mu_j|}{|\sigma_i - \sigma_j|} \big) & \frac{\mu_i-\mu_j}{\sigma_i - \sigma_j} \geq 0 \\
            2\Pr \big( Z < \frac{-\epsilon b + |\mu_i - \mu_j|}{|\sigma_i - \sigma_j|} \big) & \frac{\mu_i-\mu_j}{\sigma_i - \sigma_j} < 0
         \end{cases} \nonumber \\
    &= \begin{cases}
            2\Pr \big( Z > \frac{\epsilon b - |\mu_i - \mu_j|}{|\sigma_i - \sigma_j|} \big) & \frac{\mu_i-\mu_j}{\sigma_i - \sigma_j} \geq 0 \\
            2\Pr \big( Z > \frac{\epsilon b - |\mu_i - \mu_j|}{|\sigma_i - \sigma_j|} \big) & \frac{\mu_i-\mu_j}{\sigma_i - \sigma_j} < 0
         \end{cases} \label{eq:W1Exponential:Zsym}
\end{align}
\end{small}
where \eqref{eq:W1Exponential:Zsym} utilizes the symmetric property of standard normal distribution.
We then just need to have $\Pr \big( Z > \frac{\epsilon b - |\mu_i - \mu_j|}{|\sigma_i - \sigma_j|} \big) = \Pr \big( Z > \frac{c}{|\sigma_i - \sigma_j|} \big) \leq \frac{\delta}{2}$.

Using the $Q$-function for tail probability of standard normal distribution: $Q(t)= \frac{1}{\sqrt{2\pi}} \int_{t}^{\infty} e^{-x^2/2} \dif x$, define
\begin{equation} \label{eq:TauStar}
    \begin{aligned}
        \tau^*(\delta) &= \min \Big\{\tau \colon \Pr(Z > \tau ) \leq \frac{\delta}{2} \Big\} \\
        &= Q^{-1}(\delta/2)
    \end{aligned}
\end{equation}
We then have sufficient condition $c \geq |\sigma_i - \sigma_j| \tau^*(\delta)$ which gives
$ b \geq \frac{1}{\epsilon}  \big( |\mu_i - \mu_j| + |\sigma_i - \sigma_j| \tau^*(\delta) \big)$.
Maximizing over $\rho$ and $\SetPair$, we have \eqref{eq:theo:Lap}.

\begin{figure}[t]
	\centerline{
		\scalebox{0.7}{\input{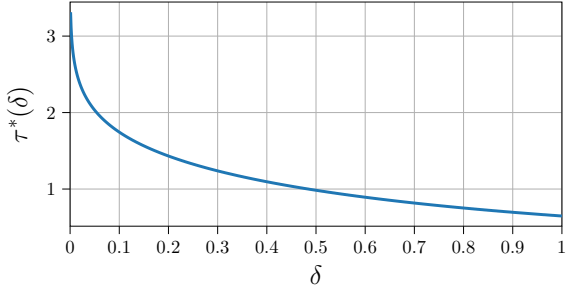}}
	}
	\caption{The value of $\tau^*(\delta)$ in \eqref{eq:TauStar1} when $\delta$ varies from $0.001$ to $0.999$.}
	\label{fig:TauStarDelta}
\end{figure}

One can derive an upper bound on $\tau^*(\delta)$ without using $Q$-function. For example, following an $(\epsilon,\delta)$-DP proof in~\cite[Appendix~A]{Dwork2014book} by using Gaussian tail bound
\begin{equation}\label{eq:GaussTail}
 \Pr(Z > \tau ) \leq \frac{1}{\sqrt{2 \pi } \tau} e^{-\frac{\tau^2}{2}}, \quad \forall \tau > 0.
\end{equation}
To ensure $\Pr(Z > \tau) \leq \frac{\delta}{2}$, it suffices to have $\frac{1}{\sqrt{2 \pi} \tau} e^{-\frac{\tau^2}{2}} \leq \frac{\delta}{2}$, which is equivalent to $\tau e^{\frac{\tau^2}{2}} \geq \frac{2}{\sqrt{2\pi}\delta} \Longrightarrow \frac{\tau^2}{2} e^{\frac{\tau^2}{2}} \geq \frac{\tau}{\sqrt{2\pi}\delta} \Longrightarrow \tau^2 \geq 2 W_0\Big( \frac{\tau}{\sqrt{2\pi}\delta}\Big)$ and
\begin{equation} \label{eq:TauStar1}
    \tau^*(\delta) \leq  \min \Big\{\tau \in \RealP \colon \tau^2 \geq 2 W_0\Big( \frac{\tau}{\sqrt{2\pi}\delta}\Big) \Big\},
\end{equation}
where $W_0(\cdot)$ is the Lambert-$W$ function such that $W_0(x)e^{W_0(x)} = x$ for nonnegative $x$. $W_0(x)$ is increasing in $x \in \RealP$. See Fig.~\ref{fig:TauStarDelta}.
Or, $\tau e^{\frac{\tau^2}{2}} \geq \frac{2}{\sqrt{2\pi}\delta} \Longrightarrow \tau^2 e^{\tau^2} \geq \frac{2}{\pi\delta^2} \Longrightarrow \tau^2 \geq W_0\Big( \frac{2}{\pi\delta^2}\Big)$ and
\begin{equation} \label{eq:TauStar2}
    \tau^*(\delta) \leq \sqrt{W_0\Big( \frac{2}{\pi\delta^2}\Big)},
\end{equation}
It should be noted that the method of determining $\tau^*(\delta)$ that satisfies $\Pr \big( Z > \frac{c}{|\sigma_i - \sigma_j|} \big) \leq \frac{\delta}{2}$ is not unique.
\hfill \IEEEQED

\section{Proof of Remark~\ref{coro:TransPrior}}
\label{app:coro:TransPrior}
For translation priors $X|s_i$ and $X|s_j$, we have linear push-forward $T(x) = \mu_j +  (x - \mu_i), \forall x \in \Real$.
Then, for Laplace noise $N \sim \Lap(b)$, we have~\eqref{eq:mainInEq} equal to
\begin{small}
\begin{align*}
     &\int \int_B \big( P_{N}(y-x) - e^{\epsilon} P_{N}(y-x -\mu_j + \mu_i) \big) \dif y \dif P_{X|S}(x|s_i,\rho)  \\
     &= \int \int_B \frac{1}{2b}  \left( e^{-\frac{|y-x|}{b}} - e^{\epsilon - \frac{|y-x -\mu_j + \mu_i|}{b}} \right) \dif y \dif P_{X|S}(x|s_i,\rho)  \\
     &\leq \big( 1 - e^{\epsilon - \frac{|\mu_i - \mu_j|}{b}} \big)  \int  \int_B \frac{1}{2b} e^{-\frac{|y-x|}{b}} \dif y \dif P_{X|S}(x|s_i,\rho), \forall B \subseteq \Real.
\end{align*}
\end{small}
Thus, any $b \geq \frac{\mu_i - \mu_j}{\epsilon}$ ensures $P_{Y|S}(y \in B|s_i,\rho) - e^\epsilon P_{Y|S}(y \in B|s_j,\rho) \leq 0 $ for all $B \subseteq \Real$. Taking the maximum of the right hand side over all $\rho$ and $(s_i,s_j) \in \SetPair$, we have~\eqref{eq:Translation}. \hfill\IEEEQED

\section{Proof of Theorem~\ref{theo:LapGaussMix}}
\label{app:theo:LapGaussMix}
For $P_{X|S}(x|s_i,\rho) = \sum_{m = 1}^{D_i} \alpha_{im} \GaussProb(x ;  \mu_{im}, \sigma_{im}^2)$ and $P_{X|S}(x|s_j,\rho) = \sum_{l = 1}^{D_j} \alpha_{jl} \GaussProb(x ;  \mu_{jl}, \sigma_{jl}^2)$ and $N \sim \Lap(b)$, we have~\eqref{eq:mainInEq} equal to
\begin{footnotesize}
		\begin{align}
		& \sum_{m} \alpha_{im}  \int \int_B \frac{1}{2b} \Big( e^{-\frac{|y-x|}{b}} - \nonumber \\
		&\qquad\qquad  e^\epsilon \sum_{l} \frac{w_{ml}^*}{\alpha_{im}}e^{-\frac{|y-T_{ml}(x)|}{b}}     \Big) \dif y \dif \GaussProb(x; \mu_{im}, \sigma_{im}^2)  \nonumber \\
		&\leq \sum_{m} \alpha_{im}  \int \int_B \frac{1}{2b} e^{-\frac{|y-x|}{b}} \dif y \Big( 1  - \nonumber \\
		&\qquad\qquad   \sum_{l} \frac{w_{ml}^*}{\alpha_{im}}e^{\epsilon-\frac{|x-T_{ml}(x)|}{b}}     \Big) \dif \GaussProb(x; \mu_{im}, \sigma_{im}^2) \nonumber \\
		&=\int \int_B \frac{1}{2b} e^{-\frac{|y-x|}{b}} \dif y \sum_{m} \alpha_{im}  \ \Big( 1 - \nonumber \\
		&\qquad\qquad \sum_{l} \frac{w_{ml}^*}{\alpha_{im}}e^{\epsilon-\frac{|x-T_{ml}(x)|}{b}}     \Big) \dif \GaussProb(x; \mu_{im}, \sigma_{im}^2) \nonumber \\
		&=\int \int_B \frac{1}{2b} e^{-\frac{|y-x|}{b}} \dif y \Big( 1 - \sum_{m,l} w_{ml}^*e^{\epsilon-\frac{|x-T_{ml}(x)|}{b}} \Big) \dif \GaussProb(x; \mu_{im}, \sigma_{im}^2)  \label{eq:ExpIneqGMM1}\\
		&\leq \int \int_B \frac{1}{2b} e^{-\frac{|y-x|}{b}} \dif y \Big( 1 - e^{\epsilon-\frac{ \sum_{m,l} w_{ml}^* |x-T_{ml}(x)|}{b}}     \Big) \dif \GaussProb(x; \mu_{im}, \sigma_{im}^2)  \label{eq:ExpIneqGMM2}\\
		&\leq \int \int_B \frac{1}{2b} e^{-\frac{|y-x|}{b}} \dif y \Big[ 1 - e^{\epsilon-\frac{ \sum_{m,l} w_{ml}^* |x-T_{ml}(x)|}{b}}     \Big]_+ \dif \GaussProb(x; \mu_{im}, \sigma_{im}^2)  \label{eq:ExpIneqGMM3}\\
		&\leq \int \Big[ 1 - e^{\epsilon-\frac{ \sum_{m,l} w_{ml}^* |x-T_{ml}(x)|}{b}}     \Big]_+ \dif \GaussProb(x; \mu_{im}, \sigma_{im}^2)  \label{eq:ExpIneqGMM4}\\
		&=\int \Big[ 1 - e^{\epsilon- \sum_{m,l} w_{ml}^* \frac{ |( \sigma_{im}-\sigma_{jl})z + (\mu_{im}-\mu_{jl}) | }{b}}     \Big]_+ \dif \GaussProb(z; 0, 1)  \label{eq:ExpIneqGMM5}\\
		&\leq \int \Big[ 1 - e^{\epsilon- \sum_{m,l} w_{ml}^* \frac{ |\sigma_{im}-\sigma_{jl}| |z| + |\mu_{im}-\mu_{jl}|  }{b}}  \Big]_+ \dif \GaussProb(z; 0, 1)  \label{eq:ExpIneqGMM6}\\
		&\leq \int_{A} \dif \GaussProb(z; 0, 1) , \qquad \forall B \subseteq \Real,  \label{eq:Aux}
	\end{align}
\end{footnotesize}
	where
$$ A = \Big\{ z \colon \sum_{m,l} w_{ml}^* \big( |\sigma_{im}-\sigma_{jl}||z| + |\mu_{im}-\mu_{jl}| \big) > \epsilon b \Big\}. $$
Equality~\eqref{eq:ExpIneqGMM1} is using the fact that $\sum_{l} w_{ml}^* = \alpha_{im}, \forall m$ and $\sum_{m} \alpha_{im} = 1$~\cite{Delon2020GMM},  \eqref{eq:ExpIneqGMM2} is using the Jensen inequality of the exponential function $\E[e^X]  \geq e^{E[X]}$,
in \eqref{eq:ExpIneqGMM3}, $[z]_+ = \max\Set{z,0}$,
inequality~\eqref{eq:ExpIneqGMM4} is because  $\int_B \frac{1}{2b} e^{-\frac{|y-x|}{b}} \dif y \leq 1$ for all $B$,
equality~\eqref{eq:ExpIneqGMM5} is by substituting $T_{ml}(x) = \mu_{jl} + \frac{\sigma_{jl}}{\sigma_{im}} (x - \mu_{im})$ and the change of variable $Z = \frac{X-\mu_{im}}{\sigma_{im}}$,
\eqref{eq:ExpIneqGMM6} is using triangular inequality  $|( \sigma_{im}-\sigma_{jl})z + (\mu_{im}-\mu_{jl}) |  \leq  |\sigma_{im}-\sigma_{jl}| |z| + |\mu_{im}-\mu_{jl}| $
and the monotonicity of $[\cdot]_+$ and  \eqref{eq:Aux} is because $\Big[ 1 - e^{\epsilon- \sum_{m,l} w_{ml}^* \frac{ |\sigma_{im}-\sigma_{jl}| |z| + |\mu_{im}-\mu_{jl}|  }{b}}  \Big]_+  \in [0,1]$.

For $b \geq \frac{1}{\epsilon} \big( \sum_{m,l} w_{ml}^*   |\mu_{im}-\mu_{jl} | + c \big)$,
\begin{align*}
    \int_{A} \dif \GaussProb(z; 0, 1) &= \Pr \Big( |Z| > \frac{c}{\sum_{m,l} w_{ml}^* |\sigma_{im}-\sigma_{jl}|} \Big) \\
    &= 2 \Pr \Big( Z > \frac{c}{\sum_{m,l} w_{ml}^* |\sigma_{im}-\sigma_{jl}|} \Big).
\end{align*}
by~\eqref{eq:GaussTail}, it suffices to have $c \geq \tau^*(\delta) \sum_{m,l} w_{ml}^* |\sigma_{im}-\sigma_{jl}|$ such that $b \geq \frac{1}{\epsilon}\sum_{m,l} w_{ml}^*  \big( |\mu_{im}-\mu_{jl} | + \tau^*(\delta)|\sigma_{im}-\sigma_{jl}| \big)$
Maximizing over $\rho$ and $\SetPair$, we get \eqref{eq:theo:LapGaussMix}. \hfill\IEEEQED

\section{Proof of Corollary~\ref{coro:TransPriorGMM}}
\label{app:coro:TransPriorGMM}
For $P_{X|S}(x|s_i,\rho) = \sum_{m = 1}^{D} \alpha_{m} \GaussProb(x ;  \mu_{im}, \sigma_{m}^2)$, $P_{X|S}(x|s_j,\rho) = \sum_{m = 1}^{D} \alpha_{m} \GaussProb(x ;  \mu_{jm}, \sigma_{m}^2)$ and $N \sim \Lap(b)$, \eqref{eq:mainInEq} equals to
\begin{footnotesize}
\begin{align*}
	 &\sum_{m} \alpha_{m} \int \int_B \Big( P_{N}(y-x)  - \\
     & \qquad\qquad  e^\epsilon P_{N}(y - x + \mu_{im} - \mu_{jm} ) \Big) \dif y \dif \GaussProb(x; \mu_{im}, \sigma_{m}^2) \\
	 & \leq \sum_{m} \alpha_{m} \Big( 1 - e^{\epsilon-\frac{|\mu_{im} - \mu_{jm}|}{b}}  \Big) \int \int_B  P_N(y-x) \dif y \dif \GaussProb(x; \mu_{im}, \sigma_{m}^2).
\end{align*}
\end{footnotesize}
So, $b \geq \frac{1}{\epsilon} \ \sum_{m} \alpha_m  |\mu_{im} - \mu_{jl}|$ ensures $\eqref{eq:mainInEq}\leq 0$. Maximizing over $\rho$ and $\SetPair$, we get \eqref{eq:TranslationGMM}. \hfill\IEEEQED

\section{Interpretation of $\hat{\pi}$}
\label{append:W2}
In this paper, the purpose of Laplace mechanism is to attain $(\epsilon,\delta)$-pufferfish privacy: for given $\epsilon$ and $\delta$,
$ P_{Y|S}(B|s_i,\rho) \leq e^\epsilon P_{Y|S}(B|s_j,\rho) + \delta , \forall B \subseteq \Real, (s_i,s_j) \in \SetPair$.
For $X|s_i \sim \Gauss(\mu_i, \sigma_i^2)$, $X|s_j\sim \Gauss(\mu_j, \sigma_j^2)$ and $N \sim \Lap(b)$,
consider the problem of searching the minimum value of $\delta$ over all couplings:
\begin{align}
	&\inf_{\pi} \{ P_{Y|S}(B|s_i,\rho) - e^\epsilon P_{Y|S}(B|s_j,\rho) \} \nonumber \\
	& = \inf_{\pi} \int_B \int \big( P_{N}(y-x) - e^\epsilon P_{N}(y-x') \big) \dif \pi(x,x') \dif y \nonumber \\
	& = \inf_{\pi} \int_B \int \frac{1}{2b} \big( e^{-\frac{|y-x|}{b}} - e^{\epsilon - \frac{|y-x'|}{b}} \big) \dif \pi(x,x') \dif y \nonumber \\
	& \leq \inf_{\pi} \int \big( 1 - e^{\epsilon - \frac{|x-x'|}{b}} \big) \int_B \frac{1}{2b} e^{-\frac{|y-x|}{b}} \dif y \dif \pi(x,x') \nonumber \\
	& \leq \inf_{\pi} \int \big[ 1 - e^{\epsilon - \frac{|x-x'|}{b}} \big]_+ \dif \pi(x,x') \nonumber \\
	& \leq \int \big[ 1 - e^{\epsilon - \frac{|x-x'|}{b}} \big]_+ \dif \hat{\pi}(x,x') \label{eq:GenIneq}  \\
	& \leq \int \big[ \frac{|x-x'|}{b}- \epsilon \big]_+  \dif \hat{\pi}(x,x') \label{eq:ExpIneq}  \\
	& \leq \left(  \int \big( \frac{|x-x'|}{b}- \epsilon \big)^2  \dif \hat{\pi}(x,x')\right)^{\frac{1}{2}} \label{eq:WMono} \\
	& = \left( \inf_{\pi} \int  \big( \frac{|x-x'|}{b}- \epsilon \big)^2  \dif \pi(x,x') \right)^{\frac{1}{2}}  , \forall B \subseteq \Real,  \label{eq:WMono1}
\end{align}
where inequality $e^x \geq 1+x$ is applied in \eqref{eq:ExpIneq} and inequality \eqref{eq:WMono} is due to the monotonicity of $\ell_\alpha$-norm.
In~\eqref{eq:WMono1}, $\inf_{\pi}$ gives rise to the Monge's transport plan $\hat{\pi}$ in \eqref{eq:WMono}, which tightens the upper bound on  $\inf_{\pi} \{ P_{Y|S}(B|s_i,\rho) - e^\epsilon P_{Y|S}(B|s_j,\rho) \}$. The reason for having this tighter upper bound is to obtain a smaller value of $b$, indicating less noise power, as a sufficient condition for $(\epsilon,\delta)$-pufferfish privacy.
Therefore, adopting Monge's optimal transport plan contributes to a reduction in noise for attaining pufferfish privacy.

In this paper, we use the Monge's solution of $\dif \hat{\pi}(x,x') = \dif P_{X|S}(x|s_i,\rho) \cdot \mathbb{I}\Set{x' = T(x)}$ and the main results, Theorems~\ref{theo:Lap} and \ref{theo:LapGaussMix}, are essentially derived by the idea of imposing an upper bound $\delta$ on \eqref{eq:GenIneq} :
\begin{equation}
		\begin{aligned}
				\int \big[ & 1 - e^{\epsilon - \frac{|x-x'|}{b}}  \big]_+ \dif \hat{\pi}(x,x')  \\
				&= \int \big[ 1 - e^{\epsilon - \frac{|x-T(x)|}{b}} \big]_+ \dif P_{X|S}(x|s_i,\rho)  \leq \delta.
		\end{aligned}
\end{equation}

\bibliography{BIB}
\bibliographystyle{IEEEtran}

\begin{IEEEbiography}[{\includegraphics[width=1in,height=1.3in,clip,keepaspectratio]{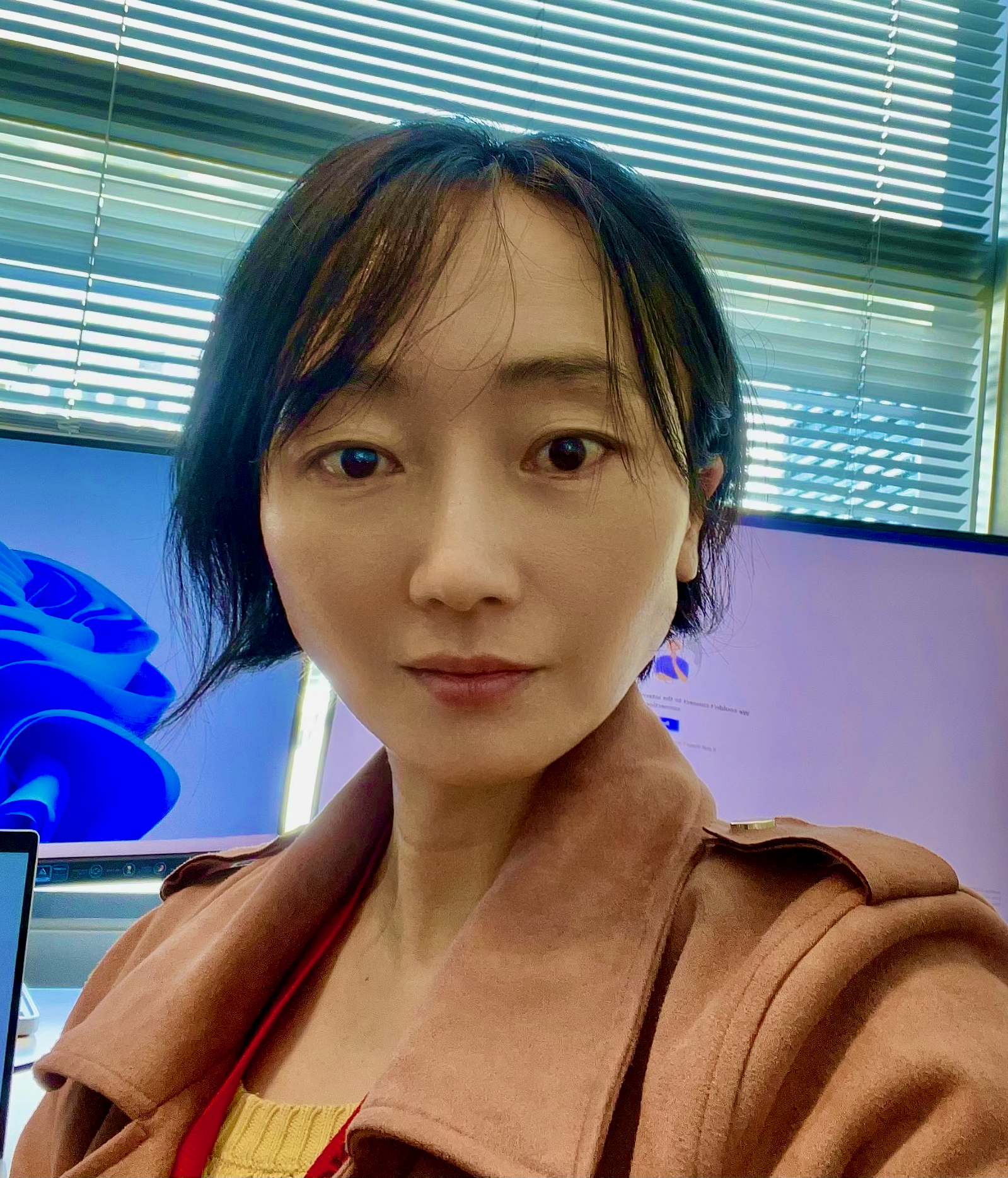}}]
	{Ni Ding}
	received the PhD degree from the Australian National University, Australia, in 2017. She was a postdoctoral fellow at Data 61, CSIRO, Australia, from 2017 to 2020 and a Doreen Thomas Postdoctoral Fellow at the University of Melbourne from 2020-2023.
	She is now a lecturer at the School of Computer Science, University of Auckland, New Zealand.
	Her research interests generally include optimizations in information theory, signal processing and machine learning. She is currently interested in data privacy, discrete and combinatorial optimization problems raised in discrete event control in cross-layer adaptive modulation, source coding and game theory (in particular, the games with strong structures, e.g., supermodular and convex games).
\end{IEEEbiography}

\end{document}